\documentclass[reprint,pre,showpacs,superscriptaddress]{revtex4-1}%nobalancelastpage
\usepackage{amsmath,amsthm,amssymb,amscd,float,graphicx}
\usepackage[utf8]{inputenc}
\usepackage[T1]{fontenc}
\usepackage{lmodern}
\usepackage[colorlinks,linkcolor=blue,citecolor=blue,urlcolor=blue]{hyperref}
\newcommand{\al}{\alpha}
\newcommand{\be}{\beta}
\newcommand{\de}{\delta}

\newcommand{\vep}{\varepsilon}
\newcommand{\ga}{\gamma}

\newcommand{\la}{\lambda}

\renewcommand{\th}{\theta}
\newcommand{\vp}{\varphi}

\newcommand{\ze}{\zeta}
\newcommand{\De}{\Delta}
\newcommand{\Ga}{\Gamma}

\newcommand{\CC}{{\mathbb C}}
\newcommand{\NN}{{\mathbb N}}
\newcommand{\RR}{{\mathbb R}}
\newcommand{\ZZ}{{\mathbb Z}}
\newcommand{\cE}{{\mathcal E}}

\newcommand{\cP}{{\mathcal P}}

\newcommand{\fH}{{\mathfrak H}}

\newcommand{\ket}[1]{|#1\rangle}
\newcommand{\bra}[1]{\langle#1|}

\newcommand{\mss}{\kern 1pt}

\renewcommand{\le}{\leqslant}
\renewcommand{\ge}{\geqslant}
\newcommand{\tends}[1]{\bbuildrel{\hbox to 2em{\rightarrowfill}}_{#1}^{}}

\newcommand{\csch}{\operatorname{csch}}

\newcommand{\sgn}{\operatorname{sgn}}
\newcommand{\Li}{\operatorname{Li}}
\newcommand{\tr}{\operatorname{tr}}

\renewcommand{\Re}{\operatorname{Re}}
\newcommand{\iu}{\mathrm i}
\newcommand{\e}{\mathrm e}
\newcommand{\diff}{\mathrm{d}}

\newcommand{\su}{\mathrm{su}}

\renewcommand{\Im}{\operatorname{Im}}

\newcommand{\en}{\enspace}

\newcommand{\Int}[1]{\,\mathop{\!#1}\limits^{\lower1ex\hbox{$\scriptstyle\circ$}}{}}
\newcommand{\Emin}{\cE_{\text{min}}}
\newcommand{\Emax}{\cE_{\text{max}}}

\theoremstyle{remark}

\begin{document}

\title{Supersymmetric spin chains with non-monotonic dispersion relation: criticality and
  entanglement entropy}

\author{José A. \surname{Carrasco}}\email{joseacar@ucm.es} \author{Federico
  \surname{Finkel}}\email{ffinkel@ucm.es} \author{Artemio
  \surname{González-López}}\email[Corresponding author. Email address: ]{artemio@ucm.es}
\author{Miguel A. \surname{Rodríguez}}\email{rodrigue@ucm.es} \affiliation{Departamento de
  Física Teórica II, Universidad Complutense de Madrid, 28040 Madrid, Spain}
\date{October 11, 2016}
\begin{abstract}
  We study the critical behavior and the ground-state entanglement of a large class of~$\su(1|1)$
  supersymmetric spin chains with a general (not necessarily monotonic) dispersion relation. We
  show that this class includes several relevant models, with both short- and long-range
  interactions of a simple form. We determine the low temperature behavior of the free energy per
  spin, and deduce that the models considered have a critical phase in the same universality class
  as a $(1+1)$-dimensional conformal field theory (CFT) with central charge equal to the number of
  connected components of the Fermi sea. We also study the Rényi entanglement entropy of the
  ground state, deriving its asymptotic behavior as the block size tends to infinity. In
  particular, we show that this entropy exhibits the logarithmic growth characteristic
  of~$(1+1)$-dimensional CFTs and one-dimensional (fermionic) critical lattice models, with a
  central charge consistent with the low-temperature behavior of the free energy. Our results
  confirm the widely believed conjecture that the critical behavior of fermionic lattice models is
  completely determined by the topology of their Fermi surface.
\end{abstract}
\maketitle

\section{Introduction}
\label{intro}

Integrable spin chains often provide a fertile ground for studying key theoretical concepts in a
simple framework that captures the essential features of the problems under consideration. An
important example of this assertion is the analysis of the entanglement of a quantum system, which
can be considered as one of the fundamental characteristics of the quantum realm~\cite{HHHH09}.
One of the most common ways of measuring the degree of entanglement of a state of a quantum
system~$X$ is via the bipartite entropy of a subsystem~$A$~\cite{NC10}. This entropy is defined
by~$S_A=S(\rho_A)$, where~$\rho_A=\tr_{X\setminus A}\rho$ is the reduced density matrix of the
subsystem~$A$, $\rho$ is the density matrix representing the state of the whole system, and $S$ is
an appropriate entropy functional (von Neumann, Rényi, etc.). The small class of models for which
the entanglement entropy can be evaluated in closed form (at least in the thermodynamic limit)
includes certain integrable spin chains, like the Lipkin--Meshkov--Glick model~\cite{LORV05}, its
$\su(n)$ generalization~\cite{CFGRT16-LMG} and the nearest-neighbors Heisenberg $X\!X$ and~$X\!Y$
models~\cite{VLRK03,JK04,IJK05}. As is well known, the latter two models are critical (gapless)
for a certain range of values of the applied magnetic field, their corresponding Virasoro algebras
having central charge respectively equal to $1$ and $1/2$. In both cases, the bipartite Rényi
entropy of a block of~$L$ consecutive spins when the whole chain is in its ground state scales as
$c(1+\al^{-1})(\log L)/6$ in the critical phase, where~$\al$ is the Rényi parameter ($\al=1$ for
the von Neumann entropy) and $c$ is the central charge. This behavior is consistent with the
scaling of the Rényi entanglement entropy of a ($1+1$)-dimensional conformal field theory
(CFT)~\cite{HLW94,CC04JSTAT,CC05}. In fact, the logarithmic scaling of the ground state
entanglement entropy is a characteristic feature of critical (fermionic) one-dimensional lattice
models with short-range interactions (see, e.g., Ref.~\cite{ECP10}).

In a previous paper~\cite{CFGRT16}, we showed that the above results also apply to a large class
of supersymmetric spin chains with general (not necessarily short-range) interactions, which turn
out to be equivalent to a suitable free fermion model. The critical character of these chains (for
appropriate values of the chemical potential~$\mu$) was ascertained via the analysis of the
low-temperature behavior of the free energy per spin. Indeed, we proved that when the dispersion
relation~$\cE(p)$ of the corresponding free fermion model is monotonic in the interval~$[0,\pi]$,
for~$0<\mu<\cE(\pi)$ the free energy per spin is approximately given (in natural
units~$\hbar=k_B=1$) by
\begin{equation}
  f(T)\simeq f_0-\frac{\pi c T^2}{6v}\,,
  \label{fTcrit}
\end{equation}
where~$v$ is the Fermi velocity (or effective speed of ``sound'') and~$c=1$. This is precisely the
expected behavior of the free energy for any critical model ($c$ being the central charge of its
Virasoro algebra), since at low temperatures the free energy of a quantum system is determined by
its lowest energy levels, and the free energy per spin of a ($1+1$)-dimensional CFT with central
charge~$c$ satisfies~\eqref{fTcrit} for sufficiently small~$T$~\cite{BCN86,Af86}. We also studied
the ground-state Rényi entanglement entropy of the above mentioned supersymmetric spin chains,
showing that in the thermodynamic limit~$L\to\infty$ it again behaves as that of a
$(1+1)$-dimensional CFT with central charge~$c=1$\,.

The aim of this paper is to extend the results of Ref.~\cite{CFGRT16} by suppressing the
requirement that the dispersion relation be monotonic in~$[0,\pi]$. As shown in
Section~\ref{sec.disp}, this makes it possible to treat a host of naturally arising models, like
supersymmetric spin chains with near and next-to-near interactions, or with long-range rational
interactions, whose dispersion relation is not always monotonic. In fact, the entanglement entropy
of fairly arbitrary energy eigenstates of one-dimensional free fermionic systems (in particular,
of the ground state of such systems with a non-monotonic dispersion relation) has been previously
studied in the literature; see, e.g., Refs.~\cite{AFC09,AEFS14}. In general, the entanglement
entropy of the ground state of these models grows logarithmically with the size~$L$ of the
subsystem, with a constant prefactor determined by the number of boundary points of the Fermi
``surface'' in $[0,2\pi)$. This logarithmic scaling is a manifestation of the so-called ``area
law'', which is believed to hold for critical fermionic systems in an arbitrary number of
dimensions~\cite{ECP10}. We shall show that the~$\su(1|1)$ supersymmetric chains studied in this
paper do indeed satisfy the area law. More precisely, by analyzing the low-temperature behavior of
the free energy we shall first show that the models under consideration are critical
for~$\Emin<\mu<\Emax$, where~$\Emin$ and~$\Emax$ respectively denote the minimum and maximum
values of the dispersion relation. (As explained in Section~\ref{sec.crit}, strictly speaking this
is only true if the roots of the equation~$\cE(p)=\mu$ are all simple.) From the latter analysis
it also follows that the central charge of these models is equal to the number~$m+1$ of disjoint
intervals that make up the Fermi sea. We shall next study the ground state Rényi entanglement
entropy, showing that in the thermodynamic limit~$L\to\infty$ it behaves as $k_\al\log L+C_\al$.
We shall explicitly compute the (non-universal) constant~$C_\al$, and prove that the
prefactor~$k_\al$ is equal to~$(m+1)(1+\al^{-1})/6$. This is in agreement with the value of the
central charge deduced from the low-temperature analysis of the free energy, and once again
confirms the conjecture that the entanglement properties of critical fermion models are entirely
determined by the topology of their Fermi surface~\cite{ECP10}.

We shall end this section with a few words on the paper's organization. In Section~\ref{sec.model}
we recall the definition of the supersymmetric chains under consideration and review their main
properties. Section~\ref{sec.disp} is devoted to the analysis of the models' dispersion relation
and the construction of simple examples of supersymmetric chains, featuring both short- and
long-range interactions, with a non-monotonic dispersion relation. In Section~\ref{sec.crit} we
derive the asymptotic behavior of the models' free energy per spin at low temperature, showing
that they are critical in an appropriate range of the chemical potential, and determine the
central charge of the corresponding Virasoro algebra. The asymptotic behavior of the entanglement
entropy of the models' ground state is determined in Section~\ref{sec.gsee} using a particular
case of the Fisher--Hartwig conjecture for Toeplitz matrices~\cite{FH68} rigorously proved
by~Böttcher and Silbermann~\cite{BS85}. We briefly state our conclusions and outline several
future developments suggested by the present work in Section~\ref{sec.conc}. The paper ends with
three appendices in which we present a review of the application of the Fisher--Hartwig conjecture
in the present context, as well as the proofs of several technical results used throughout
Section~\ref{sec.gsee}.

\section{The models}
\label{sec.model}
The type of models we shall study in this work is the class of~$\su(1|1)$ supersymmetric spin
chains with translationally invariant interactions introduced in Ref.~\cite{CFGRT16}. In the
latter models each site is occupied either by a scalar boson or a spinless fermion, whose creation
operators we shall respectively denote by~$b_i^\dagger$ and~$f_i^\dagger$, the
subindex~$i=1,\dots,N$ indicating the site on which these operators act. Thus the Hilbert space is
the $2^N$-dimensional subspace~$\fH$ of the infinite-dimensional Fock space defined by the
constraints
\begin{equation}\label{cons}
  b_i^\dagger b_i+f_i^\dagger f_i=1\,,\qquad 1\le i\le N\,.
\end{equation}
The Hamiltonian of the models under consideration is given by~\footnote{Here and in what follows,
  all sums range from $1$ to~$N$ unless otherwise stated.}
\begin{equation}\label{Hchain}
  H=\sum_{i<j}h_N(j-i)(1-S_{ij})-\mu N_f\,,
\end{equation}
where the operator $N_f$ is the total fermion number
\[
N_f=\sum_if_i^\dagger f_i\,,
\]
so that the real parameter~$\mu$ has the natural interpretation of the fermions' chemical
potential. The real-valued function~$h_N(k)$ giving the strength of the interaction between two
particles~$k$ sites apart is assumed to satisfy the constraint
\begin{equation}\label{hNconds}
  h_N(x)=h_N(N-x)\,,
\end{equation}
but is otherwise arbitrary~\footnote{Note that~$h_N(x)$ was implicitly assumed to be nonnegative
  for all~$x$ in Ref.~\cite{CFGRT16}. Here we shall drop the latter requirement, which is not
  essential in what follows.}. In other words, the chain is \emph{closed}, i.e, translationally
invariant. Finally, $S_{ij}$ is the $\su(1|1)$ spin permutation operator, defined by~\cite{Ha93}
\[
  S_{ij}=b_i^\dagger b_j^\dagger b_ib_j+f_i^\dagger f_j^\dagger f_if_j+f_j^\dagger b_i^\dagger
  f_ib_j+b_j^\dagger f_i^\dagger b_if_j\,.
\]
Equivalently, let~$\ket{s_1,\dots,s_N}\equiv\ket{s_1}\otimes\cdots\otimes\ket{s_N}$ (with
$s_k\in\{0,1\}$) be a state of the canonical spin basis, where~$\ket0$ and~$\ket1$ respectively
denote the state with one boson or one fermion. The action of~$S_{ij}$ on the latter state is then
given by
\begin{equation}\label{Sij}
  S_{ij}\ket{\dots,s_i,\dots,s_j,\dots}=(-1)^n\ket{\dots,s_j,\dots,s_i,\dots}\,,
\end{equation}
where~$n=s_i=s_j$ if $s_i=s_j$ while for $s_i\ne s_j$ $n$ equals the number of fermions in the
state~$\ket{s_1,\dots,s_N}$ occupying the sites~$i+1,\dots,j-1$. The operator~$S_{ij}$ is clearly
invariant under the supersymmetry transformation~$b_i\leftrightarrow f_i$ ($1\le i\le N$), and
on~$\fH$ we have~$N_f\mapsto N-N_f=N_b$, where $N_b=\sum_ib_i^\dagger b_i$ is the total boson
number. Hence the Hamiltonian~\eqref{Hchain} is indeed supersymmetric-invariant, up to a constant
term and the usual relabeling $\mu\mapsto-\mu$.

The fundamental feature of the $\su(1|1)$ supersymmetric chain~\eqref{Hchain}, explained in detail
in Refs.~\cite{CFGRT16,FG14JSTAT}, is that it can be mapped into a free-fermion model by
interpreting the boson state~$\ket0$ as the fermion vacuum. More precisely, consider the operators
\[
  a_i^\dagger=f^\dagger_ib_i\,,\qquad i=1,\dots,N,
\]
which can be regarded as a new set of fermion creation operators as they obviously satisfy the
canonical anticommutation relations (CAR) on $\fH$. It was shown by Haldane~\cite{Ha93} that
on~$\fH$ the $\su(1|1)$ permutation operator $S_{ij}$ can be simply expressed as
\begin{equation}
  \label{Sijak}
  S_{ij}=1-a^\dagger_ia_i-a^\dagger_ja_j+a^\dagger_ia_j+a^\dagger_ja_i\,.  
\end{equation}
Substituting into Eq.~\eqref{Hchain} we readily obtain
\begin{equation}
  \label{Hf}
  H=-\sum_{i\ne j}h_N(|i-j|)a^\dagger_ia_j-(\mu-\mu_0)\sum_ia^\dagger_ia_i,
\end{equation}
where
\[
  \mu_0=\sum_{j=1}^{N-1}h_N(j)\,.
\]
We thus see that the spin chain~\eqref{Hchain} is indeed equivalent to a free-fermion model with
hopping amplitude~$-h_N(|i-j|)$ and chemical potential~$\mu-\mu_0$.

Since the Hamiltonian~\eqref{Hf} is translationally invariant on account of Eq.~\eqref{hNconds},
it can be diagonalized by the discrete Fourier transform
\begin{equation}\label{FT}
  \hat a_l=\frac1{\sqrt N}\,\sum_{k=1}^N\e^{-2\pi\iu kl/N}a_k\,,\qquad 0\le l\le N-1\,.
\end{equation}
Indeed, the operators $\hat a_l$ obviously satisfy the CAR, and can therefore be considered as a new
set of fermionic operators. Moreover, a straightforward calculation shows that~\cite{FG14JSTAT}
\begin{equation}
  \label{Hc}
  H=\sum_{l=0}^{N-1}\Big(\vep_N(l)-\mu\Big) \hat a^\dagger_l\hat a_l\,,
\end{equation}
where
\begin{equation}\label{vepl}
  \vep_N(l)=\sum_{j=1}^{N-1}\big[1-\cos(2\pi jl/N)\big]h_N(j)\,.
\end{equation}
Likewise, the system's total momentum operator~$\cP$ is given by~\cite{BBS08}
\begin{equation*}
  \cP=\sum_{l=0}^{N-1}p_l\,\hat a^\dagger_l\hat a_l\,,
\end{equation*}
with
\[
p_l=\frac{2\pi l}N\quad (\bmod\ 2\pi)\,.
\]
Thus the operator~$\hat a^\dagger_l$ creates a (non-localized) fermion with well-defined
energy~$\vep_N(l)$ and momentum $p_l$. It follows from Eq.~\eqref{Hc} that the spectrum of~$H$ is
the set of numbers of the form
\[
  E_N(\de_0,\dots,\de_{N-1})=\sum_{l=0}^{N-1}\de_l\vep_N(l)\,,
\]
with~$\de_l\in\{0,1\}$, whose corresponding eigenstates are given by
\[
  \psi(\de_0,\dots,\de_{N-1})=(\hat a_0^\dagger)^{\de_0}\cdots(\hat a_{N-1}^\dagger)^{\de_{N-1}}\ket{0,\dots,0}\,.
\]

\section{The dispersion relation}
\label{sec.disp}

An essential requirement making it possible to study the chain~\eqref{Hchain} ---or, equivalently,
its fermionic counterpart~\eqref{Hf}--- in the thermodynamic limit is the existence of a smooth
function~$\cE(p)$ \emph{independent of $N$} such that when $N\to\infty$ we have
\begin{equation}
  \label{cEp}
  \vep_N(Np/2\pi)=\cE(p)+o(1)\,.
\end{equation}
When this is the case, we shall refer to~$\cE(p)$ as the model's \emph{dispersion relation}. From
the latter equation and the identity~$\vep_N(l)=\vep_N(N-l)$ it follows that the dispersion
relation is always symmetric about~$\pi$, namely
\begin{equation}
  \label{dispref}
  \cE(p)=\cE(2\pi-p)\,.
\end{equation}
Likewise, $\vep_N(0)=0$ implies that~$\cE(0)=0$\,. It is also customary to extend~$\cE(p)$ to the
whole real line as a~$2\pi$-periodic function, in which case Eq.~\eqref{dispref} entails
that~$\cE(p)=\cE(-p)$.

For instance, for the~$\su(1|1)$ Haldane--Shastry chain~\cite{Ha93}, whose interaction strength is
given by
\begin{equation}\label{HS}
  h_N(x)=\frac{\pi^2/N^2}{\sin^2(\pi x/N)}\,,
\end{equation}
it was shown in Ref.~\cite{GW95} that
\[
  \vep_N(l)=\frac{2\pi^2}{N^2}\,l(N-l)\,.
\]
Hence in this case~\eqref{cEp} holds with
\begin{equation}
  \cE(p)=\frac{p}2\,(2\pi-p)
  \label{dispHS}
\end{equation}
and no error term. In fact, it can be shown that Eq.~\eqref{cEp} also holds (again with no error
term) for a suitable dispersion relation~$\cE$ in the more general chain with elliptic
interactions studied in Ref.~\cite{FG14JSTAT}.

We shall next present a few relevant examples of models of the form~\eqref{Hchain} for which the
dispersion relation is guaranteed to exist. To this end, it is convenient to rewrite
Eq.~\eqref{vepl} to take into account conditions~\eqref{hNconds}, namely
\begin{multline}
  \label{veplcond}
  \vep_N(l)=2\sum_{j=1}^{\lfloor(N-1)/2\rfloor}\big[1-\cos(2\pi jl/N)\big]h_N(j)\\
  + 2[1-\pi(N)]\,\pi(l)\,h_N(N/2)\,,
\end{multline}
where~$\pi(k)\in\{0,1\}$ denotes the parity of the integer~$k$ and~$\lfloor x\rfloor$ is the
integer part of~$x\in\RR$. Clearly, the values of~$h_N(j)$ with~$1\le j\le N/2$ appearing in the
latter equation are no longer restricted by Eq.~\eqref{hNconds}. For this reason, from now on we
shall implicitly restrict the domain of~$h_N$ to the range~$1\le j\le N/2$, since
for~$N/2<j\le N-1$ we simply have~$h_N(j)=h_N(N-j)$. In this vein, we shall say (with a slight
abuse of language) that the interaction is \emph{independent of}~$N$ if there is a fixed
function~$h(x)$ such that $h_N(j)=h(j)$ for $1\le j\le N/2$\,. If this is the case we shall simply
write~$h_N=h$, again implicitly assuming that we are restricting ourselves to the
range~$1\le j\le N/2$.

An important class of models of the form~\eqref{Hchain} for which the dispersion relation~$\cE(p)$
is guaranteed to exist are those whose interaction strength~$h_N$ is short-ranged and independent
of~$N$. By this we mean that there is a positive integer~$r$ (the range of the interaction) such
that $h_N(j)=0$ for~$r<j<N-r$, and
\begin{equation}
  h_N(j)=\al_j\,,\qquad 1\le j\le r\,,
  \label{shortr}
\end{equation}
with~$\al_j$ independent of~$N$ and~$\al_r\ne0$. Obviously, in this case we have
\begin{equation}\label{cEpr}
  \cE(p)=2\sum_{j=1}^{r}\al_j\big[1-\cos(jp)\big]\,.
\end{equation}
In fact, the same is true if we drop~\eqref{shortr} but assume instead that the limit
\[
\lim_{N\to\infty}h_N(j)\equiv\al_j
\]
exists for all~$j=1,\dots,r$\,.

\begin{figure}[h]
  \centering
  \includegraphics[width=\columnwidth]{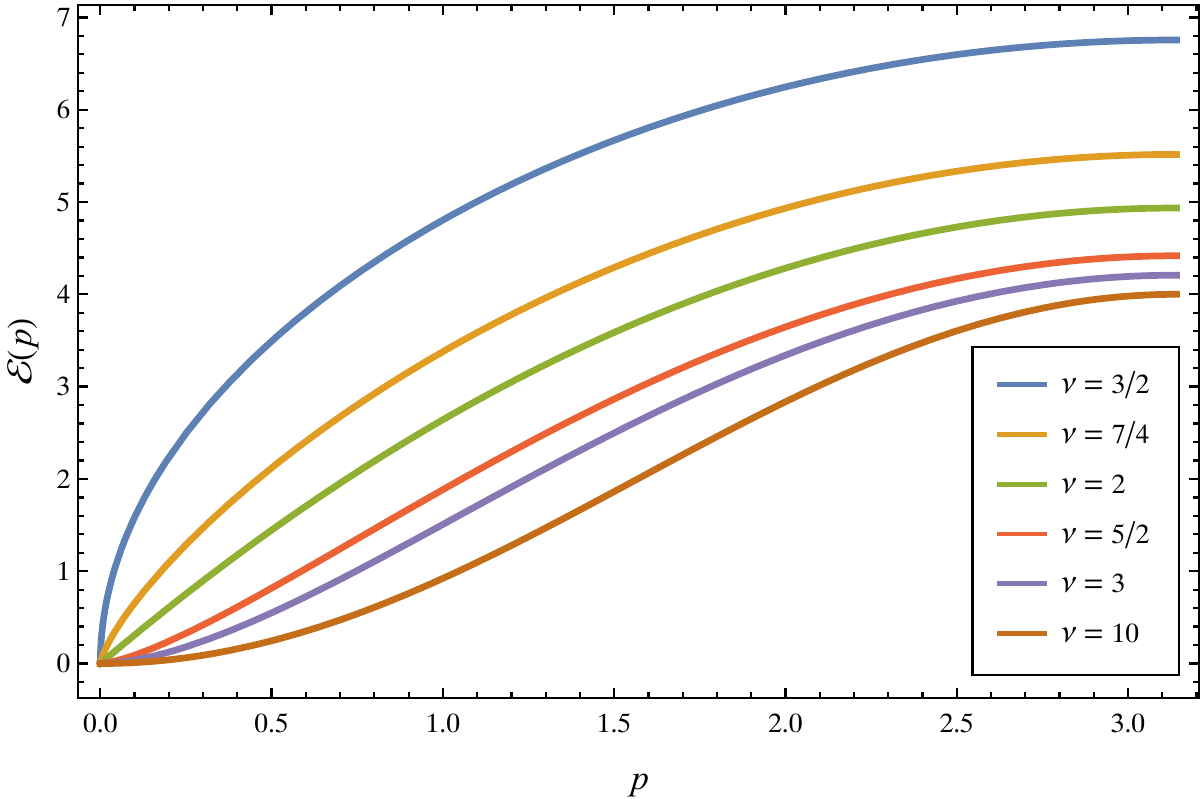}
  \caption{Dispersion relation of the~$\su(1|1)$ chain~\eqref{Hchain} with power-law
    interaction~$h_N(x)=x^{-\nu}$ for several values of the exponent~$\nu$ between $3/2$
    and~$10$.}
  \label{fig.Epp}
\end{figure}

\begin{figure}[h]
  \centering
  \includegraphics[width=\columnwidth]{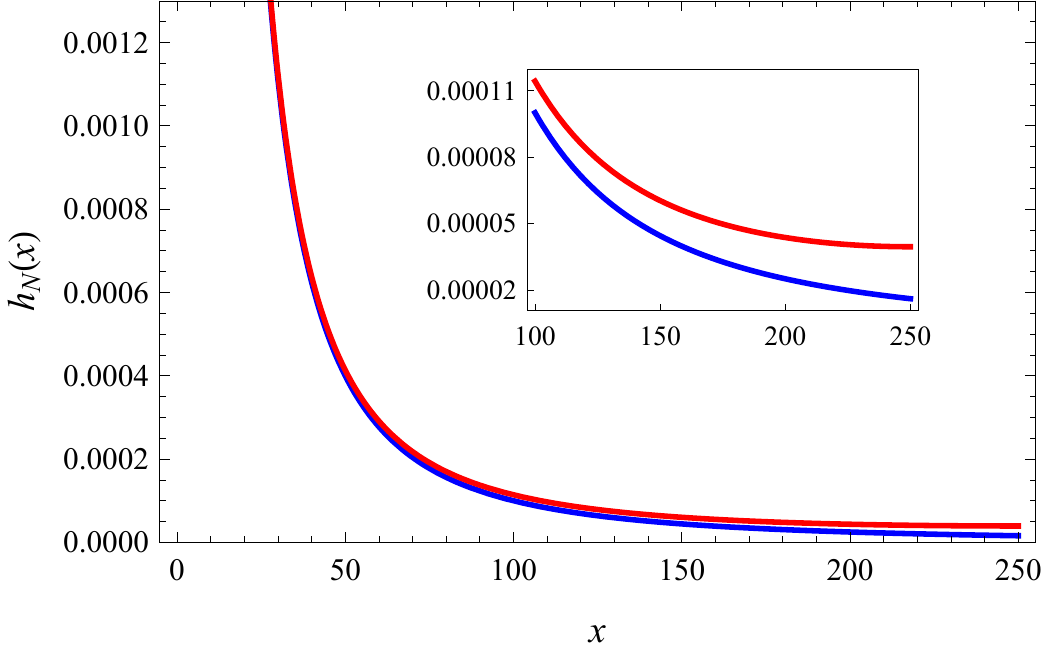}
  \caption{Comparison of the interaction strength~\eqref{HS} of the~$\su(1|1)$ HS chain (red) with
    the simple inverse-square law~$h_N(x)=1/x^2$ (blue) for~$N=500$. Inset: same plot for the
    range~$100\le x\le 250$.}
  \label{fig.HS-rat}
\end{figure}

On the other hand, the short range of the interaction~$h_N$ is by no means a necessary condition
for the existence of the dispersion relation~$\cE(p)$. Indeed, suppose for simplicity that~$h_N=h$
is independent of~$N$, and that the series~$\sum_{j=1}^\infty h(j)$ is absolutely convergent.
Then~\eqref{cEp} clearly holds with
\begin{equation}
  \label{cEpinf}
  \cE(p)=2\sum_{j=1}^\infty h(j)\big[1-\cos(jp)\big]\,.
\end{equation}
For instance, for the power law interaction~$h_N(x)=C x^{-\nu}$ with~$\nu>1$ the previous series
can be summed in closed form in terms of the polylogarithm function~\cite{OLBC10}
\[
  \Li_\nu(z)=\sum_{j=1}^\infty\frac{z^j}{j^\nu}\,,\qquad |z|\le 1\,,
\]
namely (taking, for simplicity, $C=1$)
\begin{multline}
  \cE(p)=2\ze(\nu)-\Li_\nu(\e^{\iu p})-\Li_\nu(\e^{-\iu p})\\
  =2\big[\ze(\nu)-\Re\Li_\nu(\e^{\iu
  p})\big],
  \label{cEpLi}
\end{multline}
where~$\ze$ denotes Riemann's zeta function (cf.~Fig.~\ref{fig.Epp}).
From the integral representation
\begin{equation}
  \Li_\nu(z)=\frac{z}{\Ga(\nu)}\int_0^\infty\frac{x^{\nu-1}}{\e^x-z}\,\diff x\,,
  \label{Linuint}
\end{equation}
where~$\Ga$ is Euler's gamma function, we obtain the equivalent expression
  \begin{equation}
    \label{Epint}
      \cE(p)=2\ze(\nu)-\frac2{\Ga(\nu)}\int_0^\infty\frac{(\e^x\cos p-1)x^{\nu-1}}{\e^{2x}-2\e^x\cos
    p+1}\,\diff x\,.
  \end{equation}
Using the latter formula in the identity $\cE(p)=\int_0^p\cE'(t)\diff t$ and reversing the order
of integration we arrive at the somewhat simpler expression
\[
  \cE(p)=\frac{2^\nu s}{\Ga(\nu)}\,\int_0^\infty\frac{x^{\nu-1}\coth x}{\sinh^2
    x+s}\,\diff x\,,\quad s\equiv \sin^2(p/2)\,.
\]
Remarkably, for~$\nu=2$ Eq.~\eqref{cEpLi} reduces to Eq.~\eqref{dispHS} (see, e.g.,
\cite{OLBC10}). Thus the $\su(1|1)$ chain with rational interaction~$h_N(x)=x^{-2}$ has the same
dispersion relation as the Haldane--Shastry chain~\eqref{HS}. This is of course not entirely
unexpected, since for \emph{fixed} $x\ne0$ we
have~$\lim_{N\to\infty}(\pi/N)^2\sin^{-2}(\pi x/N)=x^{-2}$. Note, however, that for~$x\sim N/2$
both interactions, although negligibly small as~$N\to\infty$, differ by a factor${}\sim(\pi/2)^2$
(cf.~Fig.~\ref{fig.HS-rat}).

Of course, although~\eqref{cEp} holds for a wide range of interesting interactions, it is not
universally true. For instance, it is not satisfied by the $N$-independent interaction
$h_N(x)=C/x$, since
\[
  \sum_{j=1}^\infty\frac{\cos(jp)}j=-\log\bigl(2\sin(p/2)\bigr)
\]
converges for $0<p<2\pi$ while the series~$\sum_{j=1}^\infty j^{-1}$ is divergent.

In a previous paper~\cite{CFGRT16} we analyzed the critical behavior of supersymmetric spin chains
of the type~\eqref{Hchain} whose dispersion relation is monotonic in the interval~$[0,\pi]$. These
models include the~$\su(1|1)$ Haldane--Shastry chain (cf.~\eqref{dispHS}) and, more generally, its
elliptic generalization introduced in Ref.~\cite{FG14JSTAT}. As is apparent
from~Fig.~\ref{fig.Epp} (and can be analytically checked differentiating Eq.~\eqref{Epint}), the
chain~\eqref{Hchain} with power-law interactions also exhibits this property. However, this
behavior is not universal, and there are in fact simple examples of supersymmetric chains of the
form~\eqref{Hchain} with a non-monotonic dispersion relation.

Indeed, consider to begin with the chain~\eqref{Hchain} with nearest and next-to-nearest
interactions, whose Hamiltonian (up to an irrelevant multiplicative constant) is given by
\begin{equation}\label{H12}
  H=\sum_i(1-S_{i,i+1})+J\sum_i(1-S_{i,i+2})-\mu N_f\,,
\end{equation}
with~$S_{N,N+1}\equiv S_{1N}$, $S_{N-1,N+1}\equiv S_{1,N-1}$ and~$S_{N,N+2}\equiv S_{2N}$\,. Note
that when~$J=0$ the fermionic version of the latter model can be mapped to the (closed) Heisenberg
$X\!X$ chain by a Wigner--Jordan transformation~\cite{CFGRT16}. From Eq.~\eqref{veplcond}
with~$h_N(1)=1$ and~$h_N(2)=J$ we easily obtain
\[
  \cE(p)=2(1-\cos p)+2J(1-\cos2p)\,.
\]
Since
\[
  \cE'(p)=2\sin p(1+4J\cos p)\,,
\]
the dispersion relation will have a critical point in~$(0,\pi)$ if and only if~$|J|>1/4$ (more
precisely, a maximum for~$J>1/4$ and a minimum for $J<-1/4$). Thus in this case the dispersion
relation is not monotonic in~$[0,\pi]$ provided that~$|J|>1/4$\,. The same is clearly true for
chains of the form~\eqref{Hchain} with interactions of finite range $r>1$, for suitable values of
the interaction strengths.

\begin{figure}[h]
  \centering
  \includegraphics[width=\columnwidth]{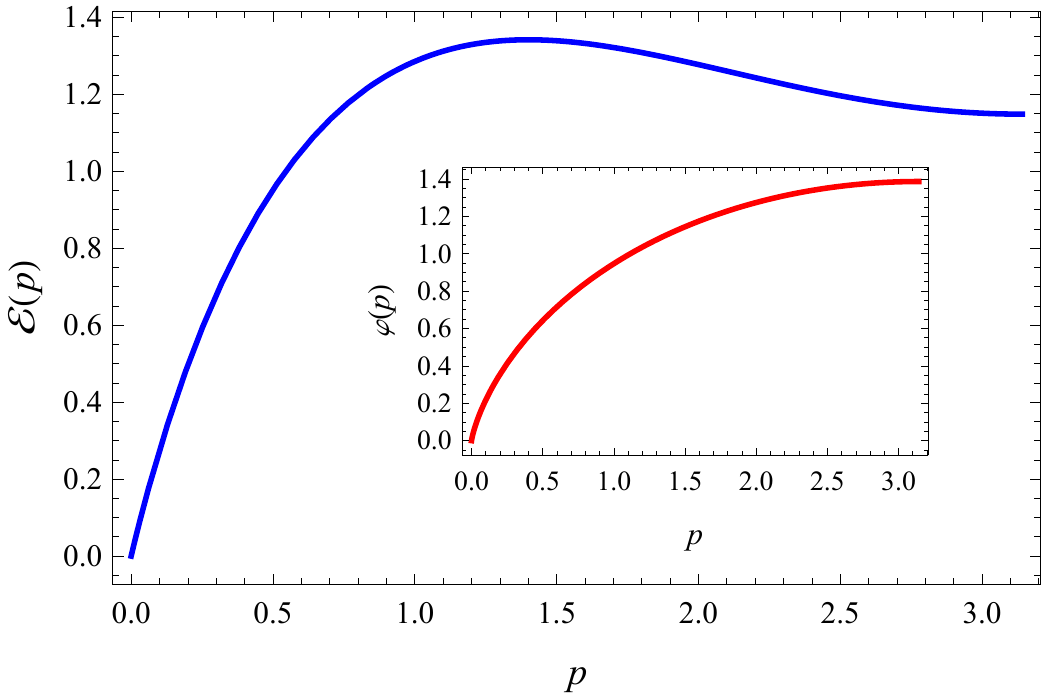}
  \caption{Dispersion relation of the spin chain~\eqref{Hchain} with interaction~\eqref{hN23}
    for~$J=0.9$. Inset: plot of the function~$\vp(p)$ in Eq.~\eqref{vpp}.}
  \label{fig.Ep23}
\end{figure}

It is also easy to construct simple examples of chains of the form~\eqref{Hchain} with long-range
interactions with a non-monotonic dispersion relation. Take, for instance,
\begin{equation}
  \label{hN23}
  h_N(x)=\frac1{x^2}-\frac{J}{x^3}\,,
\end{equation}
whose dispersion relation is given by
\[
\cE(p)=\frac p2\,(2\pi-p)-2J\big[\ze(3)-\Re\Li_3(\e^{\iu p})\big]\,.
\]
If~$p\in(0,\pi)$, differentiating the latter equation we obtain
\[
  \cE'(p)=(\pi-p)[1-J\vp(p)]\,,
\]
with
\begin{equation}
  \vp(p)=2\Im\Li_2(\e^{\iu p})/(\pi-p)\,.
  \label{vpp}
\end{equation}
It can be shown (cf.~Fig.~\ref{fig.Ep23}) that the function $\vp(p)$ increases monotonically over
the interval~$(0,\pi)$, with~$\vp(0)=0$ and $\lim_{p\to\pi-0}\vp(p)=2\log 2$~\footnote{This limit
  is easily computed from L'H\^opital's rule and the
  identity~$\Li_2'(z)=\Li_1(z)/z=-\log(1-z)/z$\,.}, so that~$\cE'(p)$ changes sign once (from
positive to negative) in~$(0,\pi)$ if and only if~$J>(2\log 2)^{-1}$. We conclude that the
dispersion relation of the chain~\eqref{Hchain} with interaction~\eqref{hN23} is not monotonic
on~$[0,\pi]$ provided that~$J>(2\log 2)^{-1}\simeq0.721348$. In particular,
for~$(2\log2)^{-1}<J<1$ the dispersion relation is not monotonic in~$[0,\pi]$ even if the
interaction strength is positive for $x\ge1$ (see~Fig.~\ref{fig.Ep23} for a plot of~$\cE(p)$
when~$J=0.9$).
\section{Critical behavior}\label{sec.crit}
In this section we shall study the critical properties of the spin chain~\eqref{Hchain} when its
dispersion relation~$\cE(p)$ is not necessarily monotonic over the interval~$[0,\pi]$. To this
end, we shall examine the low temperature behavior of the Helmholtz free energy per spin
\[
  f(T)=-T\lim_{N\to\infty}\frac{\log Z_N}N\,,
\]
which for this model is given by (cf.~\cite{CFGRT16})
\begin{equation}\label{fTchain}
  f(T)=-\frac T\pi\int_0^\pi\log\bigl[1+\e^{-\be(\cE(p)-\mu)}\bigr]\,\diff p\,.
\end{equation}
In the previous expressions~$Z_N$ denotes the partition function of the chain~\eqref{Hchain} with
$N$ spins, and $\be=1/T$ (in natural units~$\hbar=k_B=1$). As remarked in the Introduction, at low
temperatures the free energy of a critical model should satisfy Eq.~\eqref{fTcrit}. Moreover, it
was shown in Ref.~\cite{CFGRT16} that when~$\cE(p)$ is monotonic and nonnegative in the
interval~$[0,\pi]$ the model~\eqref{Hchain} is critical when the chemical potential~$\mu$ lies in
the interval~$(0,\cE(\pi))$, with central charge~$c=1$, and noncritical for~$\mu$ outside the
closed interval~$[0,\cE(\pi)]$. We shall next extend this result to the more general case in
which~$\cE(p)$ is not necessarily monotonic (nor nonnegative) in~$[0,\pi]$.

To begin with, it is immediate to show that the model~\eqref{Hchain} is not critical when~$\mu$
lies outside the interval~$[\Emin,\Emax]$. Indeed, suppose first that~$\mu<\Emin$, so that
$\cE(p)-\mu>0$, $f_0\equiv f(0)=0$ and
\[
  |f(T)|<\frac T\pi\int_0^\pi\e^{-\be(\cE(p)-\mu)}\,\diff p<T\e^{-\be(\Emin-\mu)}\,,
\]
in contradiction with the asymptotic behavior~\eqref{fTcrit} characteristic of a critical model.
Similarly, when~$\mu>\Emax$ we have
\[
  f_0\equiv f(0)=\frac1\pi\int_0^{\pi}\big[\cE(p)-\mu\big]\diff p
\]
and
\begin{align*}
  |f(T)-f_0|&=
              \frac T\pi\int_0^\pi\log\bigl[1+\e^{-\be(\mu-\cE(p))}\bigr]\,\diff p\\&<T\e^{-\be(\mu-\Emax)}\,,
\end{align*}
again in disagreement with Eq.~\eqref{fTcrit}. This conclusion is also borne out by the fact that
when~$\mu>\Emax$ or~$\mu<\Emin$ the spectrum is clearly gapped, with energy gap respectively equal
to~$\mu-\Emax$ or~$\Emin-\mu$.

Let us now consider the more interesting case in which~$\mu\in(\Emin,\Emax)$, in which the
spectrum is clearly gapless. We shall suppose that the equation~$\cE(p)=\mu$ has $m+1\ge1$
roots~$p_0<p_1<\dots<p_m$ in the interval~$(0,\pi)$, which we will assume to be \emph{simple}. We
start by expressing the free energy as
\begin{equation}\label{fTf0}
  f(T)=f_0-\frac T\pi\int_0^\pi\log\bigl(1+\e^{-\be|\cE(p)-\mu|}\bigr)\diff p\,,
\end{equation}
where
\[
  f_0\equiv f(0)=\frac1\pi\int_{\cE(p)<\mu}[\cE(p)-\mu]\diff p
\]
and the last integral is extended to the subset of the interval~$[0,\pi]$ defined by the
inequality~$\cE(p)<\mu$. Clearly, as~$T\to0+$ the main contribution to the integral in
Eq.~\eqref{fTf0} comes from an increasingly small neighborhood of the ``turning points''~$p_i$,
near which $|\cE(p)-\mu|$ is small. To exploit this fact, we choose~$\De p>0$ small enough
that~$[p_i-\De p,p_i+\De p]\cap[p_j-\De p,p_j+\De p]=\emptyset$ for $i\ne j$ and~$\cE'(p)\ne0$
on~$\cup_{i=0}^m[p_i-\De p,p_i+\De p]$. This is certainly possible, since by
hypothesis~$\cE'(p_i)\ne0$ for all~$i$. Obviously, $\De p$ depends only on the dispersion
relation~$\cE(p)$, and is therefore independent of $T$. Calling
$A=[0,\pi]-\cup_{i=0}^m[p_i-\De p,p_i+\De p]$ we have
\begin{multline}\label{fIi}
  \int_0^\pi\log\bigl[1+\e^{-\be|\cE(p)-\mu|}\bigr]\diff
  p=\int_A\log\bigl[1+\e^{-\be|\cE(p)-\mu|}\bigr]\diff p\\
  +\sum_{i=0}^m\int_{p_i-\De p}^{p_i+\De p}\log\bigl[1+\e^{-\be|\cE(p)-\mu|}\bigr]\diff p\,.
\end{multline}
The first integral can be easily estimated. Indeed, let~$a$ be the minimum value of~$|\cE(p)-\mu|$
on the compact set~$\overline A$, which is clearly positive since~$\cE(p)\ne\mu$ on~$\overline A$,
and denote by~$|A|$ the length of $A$. We then have
\begin{equation}\label{intA}
  \int_A\log\bigl[1+\e^{-\be|\cE(p)-\mu|}\bigr]\diff p\le \e^{-a\be}|A|\le\pi \e^{-a\be}\,,
\end{equation}
with $a$ (and~$|A|$) obviously independent of~$T$. Consider next the integral
\begin{equation}\label{Iidef}
  I_i\equiv\int_{p_i-\De p}^{p_i+\De p}\log\bigl[1+\e^{-\be|\cE(p)-\mu|}\bigr]\diff
  p\,.
\end{equation}
To analyze its low temperature behavior, we perform the change of variable
\begin{equation}\label{xpch}
  x=\be|\cE(p)-\mu|
\end{equation}
separately in each of the intervals~$[p_i-\De p,p_i]$ and~$[p_i,p_i+\De p]$. Since
$\cE'(p_i)\ne0$, this change of variable is one to one and~$C^\infty$ in both of the latter
intervals, and we have
\begin{multline}\label{Iix}
  \frac{I_i}{T}=\int_0^{\be|\cE(p_i-\De p)-\mu|}\log\bigl(1+\e^{-x}\bigr)\,\frac{\diff x}{|\cE'(p)|}\\
  +\int_0^{\be|\cE(p_i+\De p)-\mu|}\log\bigl(1+\e^{-x}\bigr)\,\frac{\diff x}{|\cE'(p)|}\,.
\end{multline}
The asymptotic behavior of these integrals as~$T\to0+$ can be easily determined taking into
account that by construction $\cE'(p)$ does not vanish on both intervals~$[p_i-\De p,p_i]$
and~$[p_i,p_i+\De p]$, and therefore
\[
  \frac1{\cE'(p)}=\frac{1}{\cE'(p_i)}+O(p-p_i)=\frac{1}{\cE'(p_i)}+O(Tx)\,,
\]
as~$x(p_i)=0$ implies that~$p-p_i=O(Tx)$. Since the
integral~$\int_0^\infty x\log(1+\e^{-x})\diff x$ is convergent we have
\begin{multline}\label{Iivi}
  I_i=\frac T{v_i}\bigg(\int_0^{\be|\cE(p_i-\De p)-\mu|}\log\bigl(1+\e^{-x}\bigr)\diff x\\
  +\int_0^{\be|\cE(p_i+\De p)-\mu|}\log\bigl(1+\e^{-x}\bigr)\diff x\bigg)+O(T^2)\,,
\end{multline}
where we have set
\[
  v_i=|\cE'(p_i)|\,.
\]
Moreover, if~$K$ is independent of~$\be$ we have
\begin{multline*}
  \bigg|\int_0^{K\be}\log\bigl(1+\e^{-x}\bigr)\diff x-\int_0^\infty\log\bigl(1+\e^{-x}\bigr)\diff x\bigg|
  \\
  \le\int_{K\be}^\infty\e^{-x}\diff x=\e^{-K\be}.
\end{multline*}
Using this inequality and the integral
\[
  \int_0^\infty\log\bigl(1+\e^{-x}\bigr)\diff x=\frac{\pi^2}{12}
\]
in Eq.~\eqref{Iivi} we thus have
\[
  I_i=\frac{\pi^2T}{6v_i}+O(T^2)\,.
\]
From Eqs.~\eqref{fTf0}--\eqref{Iidef} we finally obtain the asymptotic estimate
\begin{equation}\label{fasymp}
  f=f_0-\frac{\pi T^2}6\sum_{i=0}^m\frac1{v_i}+O(T^3)\,,\qquad T\to0+\,.
\end{equation}
This is the low temperature behavior of the free energy of a $(1+1)$-dimensional CFT with $m+1$
free bosons with Fermi velocities~$v_0,\dots,v_m$. Thus in this case the model~\eqref{Hchain} is
critical, with central charge~$c=m+1$.

The situation is  markedly different if any of the roots of the equation~$\cE(p)=\mu$ is not
simple. Indeed, assume that~$p_k$ is a root of order~$\nu_k>1$ of the latter equation, so that we
can write
\[
  \cE(p)-\mu=\vep_k\bigg(\frac{p-p_k}{b_k}\bigg)^{\nu_k}+O\bigl[(p-p_k)^{\nu_k+1}\bigr]
\]
with
\[
  b_k=\bigg(\frac{\nu_k!}{\big|\cE^{(\nu_k)}(p_k)\big|}\bigg)^{1/\nu_k}\,,
  \qquad\vep_k=\sgn\cE^{(\nu_k)}(p_k)\,.
\]
We now choose~$\De p>0$ such that $[p_i-\De p,p_i+\De p]\cap[p_j-\De p,p_j+\De p]=\emptyset$ for
$i\ne j$ and~$\cE'(p)\ne0$ on~$[p_i-\De p,p_i)\cup(p_i,p_i+\De p]$ for all~$i$. Proceeding as
before we again arrive at~Eqs.~\eqref{fTf0}-\eqref{fIi} and obtain the estimate~\eqref{intA} for
the first integral in Eq.~\eqref{fIi}. In order to analyze the low temperature behavior of the
integral~$I_k$, we again perform the change of variable~\eqref{xpch} in each of the
intervals~$[p_k-\De p,p_k]$ and~$[p_k,p_k+\De p]$, thus obtaining~Eq.~\eqref{Iix} with $i=k$. In
each of the latter intervals we now have
\[
  |p-p_k|=b_k(Tx)^{1/\nu_k}\big[1+O\bigl((Tx)^{1/\nu_k}\bigr)\big]
\]
and
\[
  |\cE'(p)|=\frac{\nu_k}{b_k}\bigg(\frac{|p-p_k|}{b_k}\bigg)^{\nu_k-1}\big[1+O(p-p_k)\big]\,,
\]
so that
\[
  |\cE'(p)|^{-1}=\frac{b_k}{\nu_k}\,(Tx)^{1/\nu_k-1}\big[1+O\bigl((Tx)^{1/\nu_k}\bigr)\big]\,.
\]
Substituting into Eq.~\eqref{Iix} and proceeding as before we easily obtain
\begin{align*}
  I_k&=\frac{2b_k}{\nu_k}\,T^{1/\nu_k}\!\int_0^\infty\! x^{1/\nu_k-1}\log\bigl(1+\e^{-x}\bigr)\diff x+
       O\bigl(T^{2/\nu_k}\bigr)\\
     &=2b_k\big(1-2^{-1/\nu_k}\big)\Ga\bigl(1+\nu_k^{-1}\bigr)\ze\bigl(1+\nu_k^{-1}\bigr)T^{1/\nu_k}\\
     &\hphantom{2b_k\big(1-2^{-1/\nu_k}\big)\Ga\bigl(1+\nu_k^{-1}\bigr)\ze\bigl(1+\nu_k^{-1}\bigr)}
       +O\bigl(T^{2/\nu_k}\bigr)
\end{align*}
(see Ref.~\cite{CFGRT16} for more details on the evaluation of the last integral). Thus at low
temperatures the contribution of~$p_k$ to the free energy, given by
\begin{multline}
  -\frac{TI_k}{\pi}=-\frac{2b_k}{\pi}\big(1-2^{-1/\nu_k}\big)
  \Ga\bigl(1+\nu_k^{-1}\bigr)\ze\bigl(1+\nu_k^{-1}\bigr)T^{1+1/\nu_k}\\
  +O\bigl(T^{1+2/\nu_k}\bigr)\,,
  \label{IkTpi}
\end{multline}
dominates over the~$O(T^2)$ contribution coming from the simple roots $p_i$. Moreover, since the
coefficient of~$T^{1+1/\nu_k}$ in Eq.~\eqref{IkTpi} is always negative, this term cannot be
compensated by similar terms in Eq.~\eqref{fIi} coming from other multiple roots. We thus conclude
that when~$\mu\in(\Emin,\Emax)$, but the equation~$\cE(p)=\mu$ has at least one multiple root, the
model~\eqref{Hchain} cannot be critical. A similar analysis shows that this is also the case
when~$\mu=\Emin$ or $\mu=\Emax$~\footnote{We are assuming here that~$\cE$ is smooth on~$[0,\pi]$,
  so that~$\cE'$ vanishes at the points where~$\cE$ attains its maximum and minimum values over
  the latter interval.}. This shows that the model~\eqref{Hchain} is critical if and only
if~$\Emin<\mu<\Emax$ and all the roots of the equation~$\cE(p)=\mu$ are simple. When that is the
case, the central charge of the model is equal to the number of connected components of its Fermi
sea (or, equivalently, half the number of connected components of its Fermi ``surface''). Thus,
the universality class of the model~\eqref{Hchain} depends exclusively on the topology of its
Fermi sea, which confirms the general assertion in Ref.~\cite{ECP10}.

\section{Ground state entanglement entropy}\label{sec.gsee}

As mentioned in the Introduction, one of the hallmarks of a critical fermionic lattice model in
one dimension with short-range interactions is the logarithmic growth of its ground state
bipartite entanglement entropy with the length~$L$ of the block of spins considered. More
precisely, let
\[
  S_\al=(1-\al)^{-1}\log\tr(\rho_L^\al)
\]
denote the Rényi entropy of the block when the whole chain is in its ground state~$\ket\psi$,
where~$\rho_L=\tr_{N-L}\ket\psi\bra\psi$. The expected behavior of~$S_\al$ in this type of models
is then
\begin{equation}\label{Salasymp}
  S_\al=\frac c6\,(1+\al^{-1})\log L+C_\al\,,
\end{equation}
where~$c$ is the central charge of the corresponding Virasoro algebra and~$C_\al$ is a
non-universal constant (independent of~$L$). We showed in a previous paper~\cite{CFGRT16} that the
latter formula is also valid for the supersymmetric chains~\eqref{Hchain} when their dispersion
relation is monotonic (and nonnegative) in the interval~$[0,\pi]$, even in the case of long-range
interactions. In this section we shall extend this result to a general model of the
type~\eqref{Hchain}, whose dispersion relation need not be monotonic (or nonnegative)
in~$[0,\pi]$.

\begin{figure}[h]
  \centering
  \includegraphics[width=\columnwidth]{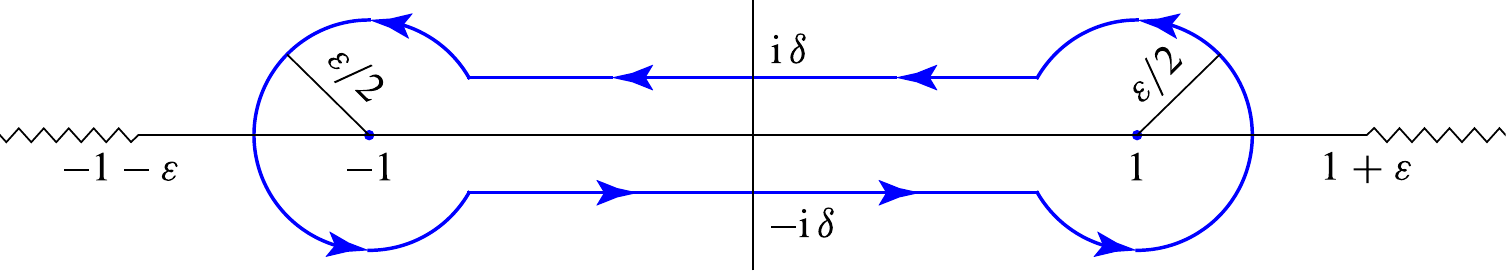}
  \caption{Integration path~$\ga_{\vep,\de}$ in Eq.~\eqref{Sintal}.}
  \label{fig.path}
\end{figure}

To this end, recall first of all that the ground state entanglement entropy~$S_\al$ can be
expressed in terms of the eigenvalues of the ground state correlation matrix~$A_L$, with matrix
elements
\[
  (A_L)_{jk}=\bra\psi a_j^\dagger a_k\ket\psi\,,\qquad 1\le j,k\le L\,.
\]
Indeed, it was shown in Ref.~\cite{VLRK03} that
\begin{equation}\label{Salla}
  S_\al=\sum_{i=1}^Ls_\al(\la_i)\,,
\end{equation}
where
\[
  s_\al(x)=(1-\al)^{-1}\log\biggl[\bigg(\frac{1+x}2\bigg)^\al +\bigg(\frac{1-x}2\bigg)^\al\biggr]
\]
and~$\la_1,\dots,\la_L\in[-1,1]$ are the eigenvalues of the matrix~$2A_L-\openone$. The asymptotic
behavior of~$S_\al$ can be determined following the method developed by Jin and
Korepin~\cite{JK04} for the $X\!X$~model. To this end, for~$\vep>0$ we define the complex-valued
function
\begin{equation}\label{salep}
  s^{(\vep)}_\al(z)=(1-\al)^{-1}\log\biggl[\bigg(\frac{1+\vep+z}2\bigg)^\al
  +\bigg(\frac{1+\vep-z}2\bigg)^\al\biggr]\,,
\end{equation}
where~$\log z\equiv\log|z|+\iu\arg_{(-\pi,\pi]}z$ and~$z^a\equiv\e^{a\log z}$. This function has a
logarithmic branch cut on the set $|\Re z|\ge1+\vep$ and no other singularities on a sufficiently
small open subset (independent of~$\vep$) containing the interval~$[-1,1]$~\footnote{Indeed,
  if~$g(z)\equiv(1+\vep+z)^\al+(1+\vep-z)^\al$ then for~$x\in[-1,1]$ $g(x)$ is real and satisfies
  $g(x)\ge2(1+\vep)^\al>2$. On the other hand, if $0<\vep<1$ then for $\ze\in\CC$ with~$|\ze|<1$
  and~$x\in[-1,1]$ integrating the derivative of~$g(z)$ along the segment from~$x$ to~$x+\ze$ we
  obtain
  \begin{align*}
    &|\Re g(x+\ze)-\Re g(x)|\\&\le\al\int_x^{x+\ze}\big|(1+\vep+z)^{\al-1}-(1+\vep-z)^{\al-1}\big|\diff
                                z\\
    &\le2\al\int_x^{x+\ze}(1+\vep+|z|)^{\al-1}\diff z\le 2\al(2+\vep+|\ze|)^{\al-1}|\ze|\\
    &\le2^{2\al-1}\al|\ze|\,.
  \end{align*} 
  Thus $\Re g(x+\ze)>0$ for $x\in[-1,1]$ provided that~$|\ze|<\min\big(1,2^{2(1-\al)}/\al\big)$.}.
By Cauchy's theorem and Eq.~\eqref{Salla}, if~$\ga_{\vep,\de}$ is the path sketched in
Fig.~\ref{fig.path} we therefore have
\begin{equation}
  \label{Sintal}
  S_\al=\lim_{\vep,\de\to0+}\frac1{2\pi\iu}
  \int_{\ga_{\vep,\de}}s^{(\vep)}_\al(\la)\frac{\diff}{\diff\la}\log D_L(\la)\,\diff\la\,,
\end{equation}
where
\begin{equation}\label{DLdef}
  D_L(\la)\equiv \det(\la+1-2A_L)\,.
\end{equation}
As explained in Appendix~\ref{app.FH}, the latter integral can then be approximated using a proved
case of the Fisher--Hartwig conjecture to estimate the logarithmic derivative of~$D_L(\la)$.

\subsection{Asymptotic formula for~$D_L(\la)$}

In order to derive the asymptotic behavior of~$D_L(\la)$, we first need to determine the symbol of
the Toeplitz matrix~$T_L\equiv\la+1-2A_L$ (see again Appendix~\ref{app.FH} for the definition of
the symbol and its calculation in two simple cases). We shall compute this symbol for a general
model of the type~\eqref{Hchain}, whose dispersion relation is not assumed to be monotonic
over~$[0,\pi]$. More precisely, we shall only suppose that the equation~$\cE(p)=\mu$ has $m+1\ge1$
\emph{simple} roots~$p_0<p_1<\dots<p_m$ in the interval~$(0,\pi)$. From the symmetry of~$\cE$
around~$\pi$ (cf.~Eq.~\eqref{dispref}) it then follows that the remaining roots of the
equation~$\cE(p)=\mu$ in the interval~$(0,2\pi)$ are $2\pi-p_m<\dots <2\pi-p_1<2\pi-p_0$.

In general, the system's ground state~$\ket\psi$ is determined by the conditions~\footnote{We are
  assuming that $p_i\not\in2\pi\ZZ/N$ for all $i=0,\dots,m$, so that the ground state is unique.}
\[
  \left\{
    \begin{aligned}
      \hat a^\dagger_k\ket\psi&=0\,,\qquad \vep_N(k)<\mu\,,\\[3pt]
      \hat a_k\ket\psi&=0\,,\qquad \vep_N(k)>\mu\,,
    \end{aligned}
  \right.
\]
so that
\[
  \bra\psi\hat a_j^\dagger\hat a_k\ket\psi=
  \begin{cases}
    0\,,\quad &\vep_N(k)>\mu\,,\\
    \de_{jk}\,, &\vep_N(k)<\mu\,.
  \end{cases}
\]
It immediately follows from Eq.~\eqref{FT} that the matrix elements of the correlation
matrix~$A_L$ are given by
\[
  (A_L)_{jk}=\frac1N\sum_{l\in I}\e^{-2\pi\iu(j-k)l/N}\,,
\]
where the sum ranges over the set $I$ of integers in the range $[0,N-1]$ satisfying the condition
$\vep_N(l)<\mu$\,. In the thermodynamic limit~$N\to\infty$ the latter formula becomes
\begin{equation}\label{ALgen}
  (A_L)_{jk}=\frac1{2\pi}\int_{\cE(p)<\mu}\e^{-\iu(j-k)p}\diff p\,,
\end{equation}
where the integral is extended to the subset of the interval~$(0,2\pi)$ defined by the
inequality~$\cE(p)<\mu$. In fact, by the~$2\pi$-periodicity of the integrand we can replace the
interval~$(0,2\pi)$ by any interval of length~$2\pi$, which we shall take as~$[-p_0,2\pi-p_0]$.
Let us suppose, for definiteness, that~$\cE'(p_0)>0$ (the case~$\cE'(p_0)<0$ is dealt with
similarly). From the simple nature of the roots~$p_j$, $2\pi-p_j$, it then follows that the
subintervals of the interval~$(-p_0,2\pi-p_0)$ on which~$\cE(p)-\mu$ is negative are
\[
  (p_{2k-1},p_{2k})\,,\qquad 0\le k\le\lfloor m/2\rfloor\,,
\]
with~$p_{-1}\equiv-p_0$, and
\[
  (2\pi-p_{2k},2\pi-p_{2k-1})\,,\qquad 1\le k\le \lfloor (m+1)/2\rfloor\,,
\]
with~$p_{m+1}\equiv2\pi-p_m$. By~Eq.~\eqref{ALgen}, the symbol of the Toeplitz
matrix~$T_L=\la+1-2A_L$ is given by
\begin{equation}\label{cgen}
  c\bigl(\e^{\iu\th}\bigr)=\begin{cases}
    \la-1\,,& \hfill-p_0<\th<p_0\,,\hfill\\
    \la+1\,,& \hfill p_0<\th<p_1\hfill\\
    \hfill\vdots\hfill&\hfill\vdots\hfill\\
    \la+1\,,& \hfill2\pi-p_2<\th<2\pi-p_1\hfill\,.
  \end{cases}
\end{equation}
Thus $c(\e^{\iu\th})$ is piecewise constant and alternates between the two values~$\la-1$ and
$\la+1$. The discontinuities of this symbol at the points~$\e^{\pm\iu p_j}$ (with $0\le j\le m$)
suggest the ansatz
\[
  c\bigl(\e^{\iu\th}\bigr)=
  b\bigl(\e^{\iu\th}\bigr)\prod_{j=0}^mt_{\be_j}\bigl(\e^{\iu(\th+p_j)}\bigr)
  t_{-\be_j}\bigl(\e^{\iu(\th-p_j)}\bigr)
\]
for suitable $b$ and~$\be_j$. To verify this ansatz, we note that for~$p_{j-1}<\th<p_j$
(with~$0\le j\le m$) we have
\begin{align*}
  t_{\be_k}\bigl(\e^{\iu(\th+p_k)}\bigr)&=\e^{\iu\be_k(\th+p_k-\pi)}\,,\\
  t_{-\be_k}\bigl(\e^{\iu(\th-p_k)}\bigr)&=
                                           \begin{cases}
                                             \e^{-\iu\be_k(\th-p_k-\pi)}\,,& 0\le k\le j-1\\
                                             \e^{-\iu\be_k(\th-p_k+\pi)}\,,& j\le k\le m\,,
                                           \end{cases}
\end{align*}
whereas for~$2\pi-p_j<\th<2\pi-p_{j-1}$ (with $1\le j\le m+1$)

\begin{align*}
  t_{\be_k}\bigl(\e^{\iu(\th+p_k)}\bigr)&=
                                          \begin{cases}
                                            \e^{\iu\be_k(\th+p_k-\pi)}\,,& 0\le k\le j-1\\
                                            \e^{\iu\be_k(\th+p_k-3\pi)}\,,& j\le k\le m\,,
                                          \end{cases}\\
  t_{-\be_k}\bigl(\e^{\iu(\th-p_k)}\bigr)&=\e^{-\iu\be_k(\th-p_k-\pi)}\,,\\
\end{align*}
and thus in either case
\[
  c\bigl(\e^{\iu\th}\bigr)=b\bigl(\e^{\iu\th}\bigr)
  \e^{2\iu\sum_{k=0}^m\be_kp_k}\e^{-2\pi\iu\sum_{k=j}^m\be_j}\,.
\]
Comparing the latter formula with Eq.~\eqref{cgen} we arrive at the system
\begin{equation}
  b\,\e^{2\iu\sum_{k=0}^m\be_kp_k}\e^{-2\pi\iu\sum_{k=j}^m\be_j}=\la-(-1)^j,\quad
  0\le j\le m+1.\label{bbeeqs}
\end{equation}
These equations easily imply that~$\be_{j}+\be_{j+1}$ is an integer multiple of~$2\pi$ for
$j=0,\dots,m-1$. We shall take $\be_j+\be_{j+1}=0$\,, so that calling~$\be_0=\be$ we have
\begin{equation}
  \label{bej}
  \be_j=(-1)^j\be\,,\qquad 0\le j\le m\,.
\end{equation}
From the equations with~$j=m$ and~$j=m+1$ we then obtain
\[
  \e^{(-1)^m2\pi\iu\be}=\frac{\la+(-1)^{m}}{\la-(-1)^m}\,,
\]
so that we can take
\begin{equation}
  \label{betagen}
  \be=\frac1{2\pi\iu}\,\log\bigg(\frac{\la+1}{\la-1}\bigg)\,.
\end{equation}
Finally, from the equation with~$j=m+1$ we have
\begin{multline}\label{bgen1}
  b=\big[\la+(-1)^{m}\big]\e^{-2\iu\be\sum_{k=0}^m(-1)^kp_k}\\
  = \big[\la+(-1)^{m}\big]\bigg(\frac{\la+1}{\la-1}\bigg)^{-\sum_{k=0}^m(-1)^k\frac{p_k}\pi}\,,
\end{multline}
which can also be written as
\begin{equation}
  \label{bgen2}
  b=\big(\la+1\big)\bigg(\frac{\la+1}{\la-1}\bigg)^{-P}\,,
\end{equation}
where
\begin{equation}
  \label{Pdef}
  P\equiv\sum_{k=0}^m(-1)^k\frac{p_k}\pi+\pi(m)
\end{equation}
and~$\pi(m)$ denotes the parity of~$m$. It is easy to check that with this choice of~$b$
and~$\be_j$ Eqs.~\eqref{bbeeqs} are all satisfied.

Since~Eq.~\eqref{betagen} coincides with the first Eq.~\eqref{bebmono}, as explained in
Appendix~\ref{app.FH}, the condition~$|\Re\be|<1/2$ is satisfied, so that we can apply the
Fisher--Hartwig conjecture to estimate $D_L(\la)\equiv\det T_L$. To this end (using the notation
in Appendix~\ref{app.FH}), note first of all that $R=2(m+1)$ and, by Eq.~\eqref{Msimp},
\begin{equation}\label{Mgen}
  M=-2(m+1)\be^2\,.
\end{equation}
Moreover, from Eq.~\eqref{bej} it easily follows that
\[
  \prod_{r=1}^RG(1+\be_r)G(1-\be_r)=\big[G(1+\be)G(1-\be)\big]^{2(m+1)}
\]
and
\begin{align*}
  \prod_{1\le s<r\le R}&\big(2|\sin\bigl(\tfrac{\th_r-\th_s}2\bigr)|\big)^{2\be_r\be_s}
                         = \prod_{i=0}^m\big(2\sin p_i\big)^{-2\be^2}\\
                       &\times\prod_{0\le j<i\le m}\big[2\sin\bigl(\tfrac{p_i-p_j}2\bigr)\big]^{4(-1)^{i+j}\be^2}\\
                       &\times 
                         \prod_{0\le j<i\le m}\big[2\sin\bigl(\tfrac{p_i+p_j}2\bigr)\big]^{-4(-1)^{i+j}\be^2}\,.
\end{align*}
Equation~\eqref{Esimp} and the Fisher--Hartwig conjecture~\eqref{detTLsimp} thus yield the
asymptotic formula
\begin{multline}
  D_L(\la)=\big[f(p_0,\dots,p_m)L^{m+1}\big]^{-2\be^2}
  (\la+1)^L\bigg(\frac{\la+1}{\la-1}\bigg)^{\!\!-LP}\\\label{DLgen} \times
  \big[G(1+\be)G(1-\be)\big]^{2(m+1)}\big(1+o(1)\big)\,,
\end{multline}
with
\begin{multline}\label{fp0pm}
  f(p_0,\dots,p_m)=\prod_{i=0}^m\big(2\sin p_i\big)\\
  \times\prod_{0\le j<i\le m}\bigg[\frac{\sin^2\bigl(\tfrac{p_i+p_j}2\bigr)}%
  {\sin^2\bigl(\tfrac{p_i-p_j}2\bigr)}\bigg]^{(-1)^{i+j}}
\end{multline}
independent of~$L$ and~$\la$.

\subsection{Asymptotic behavior of the ground state entanglement entropy}

\begin{figure}[h]
  \centering
  \includegraphics[width=.8\columnwidth]{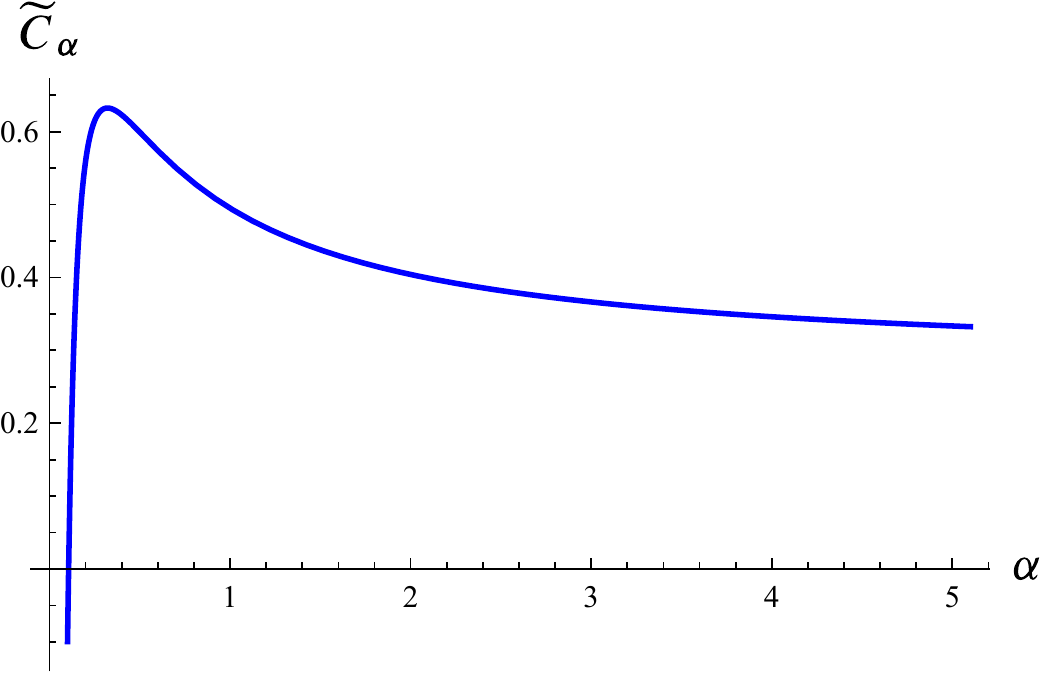}
  \caption{Plot of the constant term $\widetilde C_\al$ in Eq.~\eqref{Salfinal}.}
  \label{fig.Cal}
\end{figure}

We shall next use the approximate formula~\eqref{DLgen} and Eq.~\eqref{Sintal} to derive an
asymptotic formula for the Rényi entanglement entropy of the ground state of a general model of
the form~\eqref{Hchain} in the limit~$L\to\infty$. First of all, from Eq.~\eqref{DLgen} we easily
obtain
\begin{multline}
  \frac{\diff}{\diff\la}\log D_L(\la)\simeq L\bigg(\frac{1-P}{\la+1}+\frac{P}{\la-1}\bigg)\\
  +\frac{4\iu\be}{\pi(1-\la^2)}\Big[\log(L^{m+1}f)+(m+1)\Phi(\la)\Big]\,,
  \label{DLapprox}
\end{multline}
with
\begin{multline}\label{Phidef}
  \Phi(\la)=-\frac1{2\be}\,\frac{\diff}{\diff\be}\log\bigl[G(1+\be)G(1-\be)\bigr]\\=
  1+\ga_E+\sum_{n=1}^\infty\frac{\be^2/n}{n^2-\be^2}
\end{multline}
(cf.~Eq.~\eqref{Gsimp}).
In fact, the dominant term (proportional to~$L$) in the previous expression does not contribute to
Eq.~\eqref{Sintal}, since by Cauchy's residue theorem we have
\[
  \frac1{2\pi\iu}\int_{\ga_{\vep,\de}}s^{(\vep)}_\al(\la)\,\frac{\diff\la}{\la\mp1}=s^{(\vep)}_\al(\pm1)
  \underset{\vep\to0+}{\longrightarrow}s_\al(\pm1)=0\,.
\]
Thus Eqs.~\eqref{Sintal}-\eqref{DLapprox} yield
\begin{multline*}
  S_\al\simeq\frac2{\pi^2}\lim_{\vep,\de\to0+}\int_{\ga_{\vep,\de}}\frac{s_\al^{(\vep)}(\la)}{1-\la^2}\,
  \be\bigg[\log(fL^{m+1})\\
  +(m+1)\Phi(\la)\bigg]\diff\la\,.
\end{multline*}
Moreover, it is straightforward to verify that the integral along the circular arcs
of~$\ga_{\vep,\de}$ vanishes identically, since each of these arcs is mapped to the opposite of
the other by the transformation~$\la\mapsto-\la$, and the integrand changes sign under the latter
mapping (cf.~Eqs.~\eqref{salep}, \eqref{betagen} and~\eqref{Phidef}). We thus obtain
\begin{multline}
  \label{Slines}
  S_\al\simeq\frac2{\pi^2}\bigg(\int_{-1-\iu\,0}^{1-\iu\,0}-\int_{-1+\iu\,0}^{1+\iu\,0}\bigg)\frac{s_\al(\la)}{1-\la^2}\,
  \be\bigg[\log(fL^{m+1})\\
  +(m+1)\Phi(\la)\bigg]\diff\la\,.
\end{multline}
In order to evaluate these integrals, we note that along the segments~$\la=x\pm\iu\de$
with~$|x|<1$ we have
\[
  w\equiv\frac{\la+1}{\la-1}=\frac{x^2-1+\de^2\mp2\iu\de}{(1-x)^2+\de^2}\,,
\]
so that
\[
  \lim_{\de\to0+}|w|=\frac{1+x}{1-x}\,.
\]
On the other hand,
\[
  \Re w=\frac{x^2-1+\de^2}{(1-x)^2+\de^2}
\]
is negative for sufficiently small~$\de$, while
\[
  \frac{\Im w}{\Re w}=\frac{\pm2\de}{1-x^2-\de^2}
\]
tends to~$0$ as~$\de\to0+$ and has the same sign as~$\pm\de$, so that
\[
  \lim_{\de\to0+}\arg_{(-\pi,\pi]}w=\mp\pi\,.
\]
We thus have
\[
  \lim_{\de\to0+}\be(x\pm\iu\de)=\frac{1}{2\pi\iu}\bigg[\log\biggl(\frac{1+x}{1-x}\biggr)\mp\iu\pi\bigg]
  \equiv -\iu B(x)\mp\frac12\,,
\]
with
\[
  B(x)=\frac1{2\pi}\log\biggl(\frac{1+x}{1-x}\biggr)\,.
\]
From Eq.~\eqref{Slines} it immediately follows that
\begin{multline}\label{Sintsal}
  S_\al\simeq\big[\log\bigl(L^{m+1}f\bigr)+(m+1)(1+\ga_E)\big]I_1(\al)\\+(m+1)I_2(\al)\,,
\end{multline}
with
\begin{align}\label{I1}
  I_1(\al)&=\frac2{\pi^2}\int_{-1}^1\frac{s_\al(x)}{1-x^2}\,\diff x\,,\\
  I_2(\al)
          &=\frac4{\pi^2}\sum_{n=1}^\infty\frac1n\int_{-1}^1\frac{s_\al(x)}{1-x^2}
            \Re\bigg[\frac{\big(\frac12+\iu B(x)\big)^3}{n^2-\big(\frac12+\iu B(x)\big)^2}\bigg]
            \diff x\,.
           \label{I2}
\end{align}
The value of the integral~$I_1(\al)$ can be deduced from Ref.~\cite{JK04}
(cf.~also~\cite{AEFS14}), namely
\begin{equation}\label{I1alfinal}
  I_1(\al)=\frac{1+\al}{6\al}
\end{equation}
(see Appendix~\ref{app.int} for an elementary derivation of the latter formula). We thus obtain
\begin{equation}
  \label{Salfinal}
  S_\al\simeq(m+1)\,\frac{1+\al}{6\al}\log\bigl(Lf^{1/m+1}\bigr)+(m+1)\widetilde C_\al\,,
\end{equation}
where
\begin{multline}
  \widetilde C_\al\equiv \frac{1+\al}{6\al}\,(1+\ga_E)+I_2(\al)=\frac{1+\al}{6\al}\,(1+\ga_E)\\
  +\frac4{\pi^2}\sum_{n=1}^\infty\frac1n\int_{-1}^1\frac{s_\al(x)}{1-x^2}
  \Re\bigg[\frac{\big(\frac12+\iu B(x)\big)^3}{n^2-\big(\frac12+\iu B(x)\big)^2}\bigg] \diff x\,.
  \label{Calfinal}
\end{multline}
Comparing with Eq.~\eqref{Salasymp}, we see that the ground-state Rényi entanglement entropy of
the model~\eqref{Hchain} behaves as that of a critical system with central charge~$c=m+1$, as
expected. Moreover, the constant term~$C_\al$ is given in this case by
\begin{equation}\label{CalCtal}
C_\al=\frac{1+\al}{6\al}\,\log f(p_0,\dots,p_m)+(m+1)\widetilde C_\al\,,
\end{equation}
where the first term is model dependent (it depends on~$\mu$ and~$\cE(p)$ through the
momenta~$p_i$), while~$\widetilde C_\al$ is a universal constant (independent of~$L$ and $p_i$)
characteristic of the class of models under consideration. It is shown in Appendix~\ref{app.Cal}
that~$\widetilde C_\al$ in Eq.~\eqref{Salfinal} can be expressed as
\begin{equation}\label{Calpsi}
  \widetilde C_\al=-\frac2{\pi^2}\int_{-1}^1\frac{s_\al(x)}{1-x^2}\,\Re\big[\psi\bigl(\tfrac12+\iu
  B(x)\bigr)\big]\diff x\,,
\end{equation}
where
\[
  \psi(z)=\frac{\diff}{\diff z}\log\Gamma (z)
\]
is the digamma function. In particular, Eq.~\eqref{Calpsi} implies that~$\widetilde C_\al$
coincides with the function~$\varUpsilon_1^{(\al)}$ defined in Eq.~(64) of Ref.~\cite{JK04}. Since
for $m=0$ we have $f(p_0)=2\sin p_0$, Eq.~\eqref{Salfinal} yields the formula derived in
Ref.~\cite{CFGRT16} for the Rényi entanglement entropy of the model~\eqref{Hchain} when its
dispersion relation is monotonic over the interval~$[0,\pi]$ (which, as explained in the latter
reference, includes the~$X\!X$ model studied in Ref.~\cite{JK04}). In fact, using the ideas of
Ref.~\cite{JK04} Eq.~\eqref{Calpsi} can be written in the simpler form
\begin{multline}
  \label{Calsimp}
  \widetilde C_\al=\frac1{1-\al}\int_0^\infty\bigg(\al\csch^2t
  -\csch t\csch(t/\al)\\-\frac{1-\al^2}{6\al}\,\e^{-2t}\bigg)\frac{\diff t}t
\end{multline}
(see Appendix~\ref{app.Cal} for details). From the previous expression it is straightforward to
evaluate~$\widetilde C_\al$ numerically for any specific value of the Rényi parameter~$\al>0$;
cf.~Fig.~\ref{fig.Cal}.
 It can be numerically verified that $\widetilde C_\al$
vanishes for~$\al\simeq0.106022$, and attains its maximum value $\simeq0.632417$
for~$\al\simeq0.321699$ (cf.~Fig.~\ref{fig.Cal}).
It also follows from Eq.~\eqref{Calsimp} that $\widetilde C_\al\to-\infty$ as $\al\to0+$, and that
when~$\al\to\infty$ $\widetilde C_\al$ tends to a finite (nonzero) limit, given by
\[
  \widetilde C_\infty=\int_0^\infty\bigg(\frac1t\,\csch t-\csch^2t-\frac{\e^{-2t}}{6}\bigg)\frac{\diff
    t}{t}\simeq 0.27970\,.
\]

Taking the~$\al\to1$ limit in the previous formulas we obtain the following asymptotic expression
for the von Neumann entropy~$S\equiv S_1$:
\begin{equation}\label{SVN}
  S\simeq\frac{m+1}3\,\log L+C_1\,,
\end{equation}
where
\[
  C_1=\frac13\log f(p_0,\dots,p_m)+(m+1)\widetilde C_1
\]
and the universal constant~$\widetilde C_1\equiv\lim_{\al\to 1}\widetilde C_\al$ is given by
\[
  \widetilde C_1=\int_0^\infty\bigg(\frac{\cosh t}{\sinh^3t}-\frac1{t\sinh^2t}-
  \frac{\e^{-2t}}{3t}\bigg)\diff t\simeq0.495018\,.
\]
Note, in particular, that the latter equation agrees with the formula for the analogous
constant~$\varUpsilon_1$ in Ref.~\cite{JK04}.

\begin{figure}[h]
  \centering
  \includegraphics[width=\columnwidth]{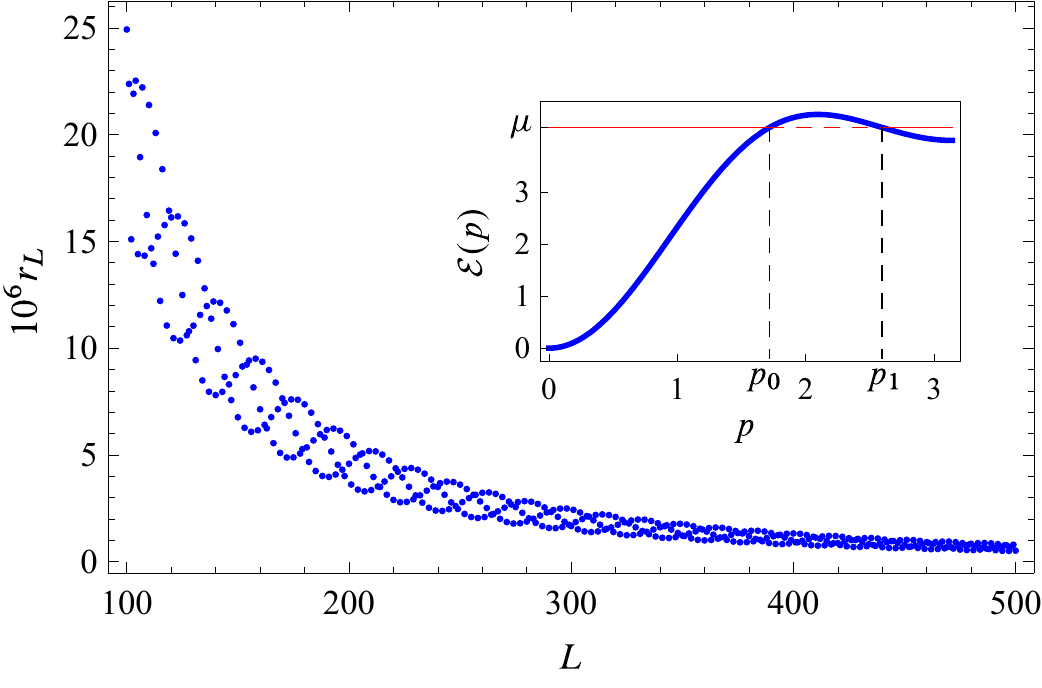}
  \caption{Relative error~$r_L\equiv S_{\mathrm{app}}/S-1$ of~the approximation $S_{\mathrm{app}}$
    in the RHS of~Eq.~\eqref{SVN} to the von Neumann ground-state entanglement entropy of the
    chain~\eqref{H12} with $J=1/2$ and~$\mu=17/4$ as a function of the block length~$L$. Inset:
    dispersion relation of the chain~\eqref{H12} with the latter values of~$J$ and~$\mu$. (The
    interval $(\cE(\pi),\Emax)$ is in this case is~$(4,9/2)$, $m=1$, $p_0\simeq1.71777$
    and~$p_1\simeq2.59356$.)}
  \label{fig.error}
\end{figure}

The formula~\eqref{Salfinal}-\eqref{Calsimp} (or its counterpart~\eqref{SVN} for the von Neumann
entropy) provides an excellent approximation to the ground-state Rényi entanglement entropy of the
supersymmetric chain~\eqref{Hchain} for even moderately large values of~$L$. As an example, in
Fig.~\ref{fig.error} we have represented the relative error $r_L\equiv S_{\mathrm{app}}/S-1$,
where~$S_{\mathrm{app}}$ is the approximation~\eqref{SVN} to the von Neumann entropy~$S$, for the
finite-range chain~\eqref{H12} in the case~$J=1/2$ and~$\mu=17/4$. The value of~$S$ has been
numerically computed diagonalizing the correlation matrix~$A_L$ and using the exact
formula~\eqref{Salla} (with~$\al=1$). As explained in Section~\ref{sec.disp}, for the value of~$J$
considered the dispersion relation has exactly one maximum in the interval~$(0,\pi)$, and hence is
not monotonic. In particular, for~$\mu\in(\cE(\pi),\Emax)=(4,9/2)$ the Fermi sea consists of two
disjoint intervals, as is also apparent from the inset in~Fig.~\ref{fig.error}. As can be seen
from the latter figure, the relative error decreases (though not monotonically)
from~$2.5\times 10^{-5}$ to $10^{-6}$ when $L$ ranges from~$100$ to~$500$.

\section{Conclusions and outlook}\label{sec.conc}

In this paper we have analyzed the critical behavior of a large class of supersymmetric spin
chains whose dispersion relation~$\cE(p)$ is not assumed to be monotonic in the
interval~$[0,\pi]$. We have examined the conditions under which the dispersion relation is
well-defined (i.e., is a continuous function) in the thermodynamic limit, providing several simple
examples of models of this type, with both short- and long-range interactions, whose dispersion
relation is not monotonic.

The main conclusion of our work is that the criticality properties of the supersymmetric
chains~\eqref{Hchain} are determined exclusively by the topology (the number of points) of their
Fermi ``surface''. More precisely, through the analysis of the free energy per spin in the
critical (gapless) phase, we have shown that these models are equivalent to a system of $m+1$ free
bosons with Fermi velocities $v_i=\cE'(p_i)$, where~$p_0,\dots,p_m$ are the points of the Fermi
surface in the interval~$[0,\pi]$. In particular, the central charge is equal to the number~$m+1$
of connected components (intervals) of the Fermi sea. This result is corroborated by the
asymptotic behavior of the ground-state Rényi entanglement entropy~$S_\al$ as the block size~$L$
tends to infinity, which has been derived applying a proved case of the Fisher--Hartwig
conjecture. Indeed, we have shown that $S_\al\simeq(m+1)(1+\al^{-1})\log L+C_\al$, where $C_\al$
is a nonuniversal constant (independent of~$L$) which we have computed in closed form in terms of
the momenta~$p_0,\dots,p_m$. In particular, for large~$L$ the entanglement entropy exhibits the
logarithmic growth characteristic of $(1+1)$-dimensional conformal field theories with central
charge~$c=m+1$. This behavior, which is typical of critical (fermionic) one-dimensional lattice
models with short-range interactions (see, e.g.,~\cite{ECP10,AEFS14}), was recently established by
the authors for supersymmetric spin chains of the type considered here under the assumption that
the dispersion relation is monotonic in~$[0,\pi]$.

The present work opens up several possible lines for future research. In the first place, one
could consider a generalization of our results on the ground-state entanglement entropy to more
general situations (for instance, considering excited states, as in Ref.~\cite{AEFS14}), in which
the Fermi sea is not necessarily a finite union of disjoint intervals but exhibits a more
complicated topological structure. Another interesting generalization of the present work is the
analysis of the entanglement of a subset consisting of the union of two or more disjoint blocks.
In fact, the entanglement entropy of this type of subsystems has already been discussed in
Ref.~\cite{AEF14}, giving rise to an unproved conjecture on the asymptotic behavior of the
determinant of a block Toeplitz matrix.

\section*{Acknowledgments}
The authors would like to thank P. Tempesta for several enlightening discussions. This work was
partially supported by Spain's MINECO under research grant no.~FIS2015-63966-P. JAC would also
like to acknowledge the financial support of the Universidad Complutense de Madrid through a 2015
predoctoral scholarship.

\appendix
\section{Toeplitz matrices and the Fisher--Hartwig conjecture}\label{app.FH}

In this appendix we shall briefly review the Fisher--Hartwig conjecture on the asymptotic behavior
of the determinant of a Toeplitz matrix when its order tends to infinity. (Recall that a
matrix~$T$ is Toeplitz if its matrix elements~$t_{ij}$ depend only on~$i-j$.)

If $c(z)$ is a (complex-valued) function defined on the unit circle~$S^1=\{z\in\CC:|z|=1\}$\,, we
define its Fourier coefficients~$c_n$ ($n\in\ZZ$) by
\[
  c_n=\frac1{2\pi\iu}\int_{|z|=1}c(z)\,z^{-n-1}\,\diff
  z\equiv\frac1{2\pi}\int_0^{2\pi}c\bigl(\e^{\iu\th}\bigr)\e^{-\iu n\th}\,\diff\th\,.
\]
Note that the last integral can in fact be extended to any interval of length~$2\pi$, by the
$2\pi$-periodicity of the integrand. For any~$L\in\NN$, the function~$c:S^1\to\CC$ defines a
Toeplitz matrix~$T_L$ of order~$L$ through the relation
\[
  (T_L)_{ij}=c_{i-j}\,,\qquad 1\le i,j\le L\,.
\]
We shall say that the function~$c$ is the \emph{symbol} of the Toeplitz matrix~$T_L$. The
Fisher--Hartwig conjecture applies to matrices $T_L$ whose symbol satisfies certain requirements
that we shall next describe.

More precisely~\cite{ES97}, $c$ should be of the form
\begin{equation}
  c(z)=b(z)\prod_{r=1}^Rt_{\be_r}\bigl(\e^{\iu(\th-\th_r)}\bigr)\big[2-2\cos(\th-\th_r)\big]^{\al_r}\,,
  \label{cz}
\end{equation}
where~$\Re\al_r>-1/2$, $b:S^1\to\CC$ is a nonvanishing smooth function with zero winding number
and
\begin{equation}\label{tbe}
  t_\be(z)=\e^{\iu\be(\th-\pi)}\,,\qquad \th\equiv\arg_{[0,2\pi)}z\,.
\end{equation}
Note that $t_\be(\e^{\iu(\th-\th_0})$ has in general (i.e., unless $\be$ is an integer) a single
jump discontinuity at~$z=\e^{\iu\th_0}$. If $c$ satisfies Eq.~\eqref{cz}, we denote by~$l_n$
($n\in\ZZ$) the~$n$-th Fourier coefficient of~$\log b$ (which is well defined and smooth, from the
smoothness of $b$ and the assumption on its winding number), and define
\[
  b_\pm(z)=\exp\bigg(\sum_{n=1}^\infty l_{\pm n}z^{\pm n}\bigg)\,,\qquad z\in S_1\,.
\]
It is immediate to show that $b_+$ (resp.~$b_-$) can be analytically prolonged to the interior
(resp.~exterior) of the unit circle. It also follows from the definition of~$b_{\pm}$ that on the
unit circle we have the \emph{Wiener--Hopf decomposition}
\[
  b(z)=\e^{l_0}b_+(z)b_-(z)\,,\qquad z\in S^1\,.
\]
Let us further set
\[
  E[b]=\exp\bigg(\sum_{n=1}^{\infty}n\,l_nl_{-n}\bigg)
\]
and
\begin{align*}
  E&=E[b]\prod_{r=1}^Rb_+\bigl(\e^{\iu\th_r}\bigr)^{\be_r-\al_r}b_-\bigl(\e^{\iu\th_r}\bigr)^{-\al_r-\be_r}\\
   &\hphantom{={}}\times\prod_{1\le s\ne r\le R}\big(1-\e^{\iu(\th_s-\th_r)}\big)^{(\al_r+\be_r)(\be_s-\al_s)}\\
   &\hphantom{={}}\times\prod_{r=1}^R\frac{G(1+\al_r+\be_r)G(1+\al_r-\be_r)}{G(1+2\al_r)}\,,
\end{align*}
where the Barnes $G$-function is the entire function defined by
\[
  G(1+z)=(2\pi)^{\frac z2}\e^{-(z+1)\frac
    z2-\ga_E\frac{z^2}2}\prod_{n=1}^\infty\bigg[\bigg(1+\frac
  zn\bigg)^{\!n}\e^{-z+\frac{z^2}{2n}}\bigg]
\]
and~$\ga_E$ is the Euler--Mascheroni constant. The Fisher--Hartwig conjecture states~\cite{ES97}
that if~$T_L$ is the Toeplitz matrix with symbol~\eqref{cz} then when $L\to\infty$ we have
\[
  \det T_L=\e^{l_0L}L^ME\big(1+o(1)\big)\,,
\]
with
\[
  M=\sum_{r=1}^R(\al_r^2-\be_r^2)\,.
\]
The above conjecture was actually proved by Böttcher and Silbermann~\cite{BS85} in the case
\[
  |\Re\al_r|<\frac12\,,\quad |\Re\be_r|<\frac12\,,\qquad r=1,\dots,R\,,
\]
which, as we shall see below, is the relevant one for our purposes. In fact, as explained in
Section~\ref{sec.gsee}, we shall only need to consider the case in which $\al_r=0$ for all $r$
and~$b$ is a constant (i.e., independent of~$\th$). The Fisher--Hartwig conjecture simplifies
considerably in this case, since
\[
  l_n=l_0\,\de_{0n}\implies b_\pm=E[b]=1\,,\quad\e^{l_0}=b\,,
\]
and therefore
\begin{equation}\label{detTLsimp}
  \det T_L=b^LL^ME\big(1+o(1)\big)
\end{equation}    
with
\begin{equation}
  M=-\sum_{r=1}^R\be_r^2
  \label{Msimp}
\end{equation}
and
\begin{multline}\label{Esimp}
  E=\prod_{1\le s<r\le R}\left(2\left|\sin\bigl(\tfrac{\th_r-\th_s}2\bigr)\right|\right)^{2\be_r\be_s}\\
\times\prod_{r=1}^RG(1+\be_r)G(1-\be_r)\,.
\end{multline}
 Note also that the
product~$G(1+z)G(1-z)$ reduces to
\begin{equation}
  G(1+z)G(1-z)=\e^{-(1+\ga_E)z^2}
  \prod_{n=1}^\infty\bigg[\bigg(1-\frac{z^2}{n^2}\bigg)^n\e^{\frac{z^2}n}\bigg]\,.
  \label{Gsimp}
\end{equation}

We shall be mainly interested in the case in which~$T_L=\la-(2A_L-\openone)$, where~$\la$ is a
spectral parameter and $A_L$ is the correlation matrix of a block of $L$ spins of the~$\su(1|1)$
supersymmetric spin chain. As a first example, let us express in the form~\eqref{cz}-\eqref{tbe}
the symbol of the matrix~$T_L$ when the chain's dispersion relation is monotonic in the
interval~$[0,\pi]$. To begin with, in this case we have
\[
  (A_L)_{jk}=\frac{\sin(p_0(j-k))}{\pi(j-k)}=\frac1{2\pi}\int_{-p_0}^{p_0}\e^{-\iu(j-k)\th}\,\diff\th,
\]
where~$p_0\in[0,\pi]$ is the Fermi momentum~\cite{CFGRT16}. Thus the symbol of the Toeplitz
matrix~$A_L$ is
\[
  f\bigl(\e^{\iu\th}\bigr)=
  \begin{cases}
    1\,,\quad& -p_0<\th<p_0\,,\\
    0\,,& p_0<\th<2\pi-p_0\,,
  \end{cases}
\]
and that of~$T_L$ is therefore given by
\[
  c\bigl(\e^{\iu\th}\bigr)=\begin{cases}
    \la-1\,,\quad& -p_0<\th<p_0\,,\\
    \la+1\,,& p_0<\th<2\pi-p_0\,.
  \end{cases}
\]
Note that~$c$ has two jump discontinuities on the unit circle at the points~$\e^{\pm\iu p_0}$. We
shall next show that
\begin{equation}\label{cbt}
  c\bigl(\e^{\iu\th}\bigr)=b\bigl(\e^{\iu\th}\bigr)t_{\be}\bigl(\e^{\iu(\th+p_0)}\bigr)t_{-\be}\bigl(\e^{\iu(\th-p_0)}\bigr)
\end{equation}
for suitable~$\be$ and $b(z)$. Indeed, first of all we have
\begin{align*}
  -p_0<\th<2\pi-p_0
  &\implies 0<\th+p_0<2\pi\\
  &\implies t_{\be}\bigl(\e^{\iu(\th+p_0)}\bigr)=\e^{\iu\be(\th+p_0-\pi)}\,.
\end{align*}
On the other hand, if~$-p_0<\th<p_0$ then
\begin{multline*}
  0\le2(\pi-p_0)<\th-p_0+2\pi<2\pi\\
  \implies t_{-\be}\bigl(\e^{\iu(\th-p_0)}\bigr)=\e^{-\iu\be(\th-p_0+\pi)}\,,
\end{multline*}
while for~$p_0<\th<2\pi-p_0$ we have
\begin{multline*}
  0<\th-p_0<2(\pi-p_0)\le2\pi\\
  \implies t_{-\be}\bigl(\e^{\iu(\th-p_0)}\bigr)=\e^{-\iu\be(\th-p_0-\pi)}\,.
\end{multline*}
Hence
\begin{equation}
  \begin{aligned}
    t_{\be}\bigl(\e^{\iu(\th+p_0)}\bigr)&t_{-\be}\bigl(\e^{\iu(\th-p_0)}\bigr)\\
    &=
    \begin{cases}
      \e^{2\iu\be(p_0-\pi)}\,,& -p_0<\th<p_0\,,\\
      \e^{2\iu\be p_0}\,,& p_0<\th<2\pi-p_0\,.
    \end{cases}
  \end{aligned}\label{tbp0}
\end{equation}
In order for Eq.~\eqref{cbt} to hold we must therefore have
\[
  b\,\e^{2\iu\be(p_0-\pi)}=\la-1\,,\qquad b\,\e^{2\iu\be p_0}=\la+1\,,
\]
from which we easily get
\[
  \e^{2\iu\be\pi}=\frac{\la+1}{\la-1}\,,\qquad b=(\la+1)\e^{-2\iu\be p_0}\,.
\]
Although these equations admit an infinite number of solutions~$(\be,b)$ provided
that~$\la\ne\pm1$, it will prove convenient for our purposes to take
\begin{equation}\label{bebmono}
  \be=\frac1{2\pi\iu}\,\log\bigg(\frac{\la+1}{\la-1}\bigg)\,,\quad
  b=(\la+1)\bigg(\frac{\la+1}{\la-1}\bigg)^{-p_0/\pi}\,,
\end{equation}
where $\log z\equiv\log|z|+\iu\arg_{(-\pi,\pi]}z$ and $z^a\equiv\e^{a\log z}$. Note, in
particular, that $b$ is a nonvanishing constant. It is also important to observe that
\[
  \la\notin[-1,1]\implies|\Re\be|=\frac1{2\pi}\arg_{(-\pi,\pi]}\bigg(\frac{\la+1}{\la-1}\bigg)<\frac12\,,
\]
since by definition $-\pi<\arg_{(-\pi,\pi]}z\le\pi$ and
\begin{align*}
  \arg_{(-\pi,\pi]}\bigg(\frac{\la+1}{\la-1}\bigg)=\pi&\iff
                                                        \frac{\la+1}{\la-1}\in(-\infty,0)\\&\iff\la\in(-1,1)\,.
\end{align*}
Thus the Fisher--Hartwig conjecture can be applied provided that~$\la$ lies outside the closed
interval~$[-1,1]$, with~$R=1$, $\al_1=0$ and
\[
  M=-2\be^2\,,\qquad E=\big(2\sin p_0\big)^{-2\be^2}G(1+\be)^2G(1-\be)^2\,.
\]
By Eqs.~\eqref{detTLsimp} and~\eqref{bebmono}, when~$L\to\infty$ the characteristic polynomial
$D_L(\la)\equiv \det(\la+1-2A_L)$ is given by
\begin{multline}\label{DLp0}
  D_L(\la)=\big(2L\sin p_0\big)^{-2\be^2}
  (\la+1)^L\bigg(\frac{\la+1}{\la-1}\bigg)^{-Lp_0/\pi}\\\times
  G(1+\be)^2G(1-\be)^2\big(1+o(1)\big)\,,
\end{multline}
with $\be$ defined by Eq.~\eqref{bebmono}. This is precisely the formula used by Jin and
Korepin~\cite{JK04} for the determination of the asymptotic behavior of the ground state
entanglement entropy of the $X\!X$ model.

\begin{figure}[h]
  \centering
  \includegraphics[width=.8\columnwidth]{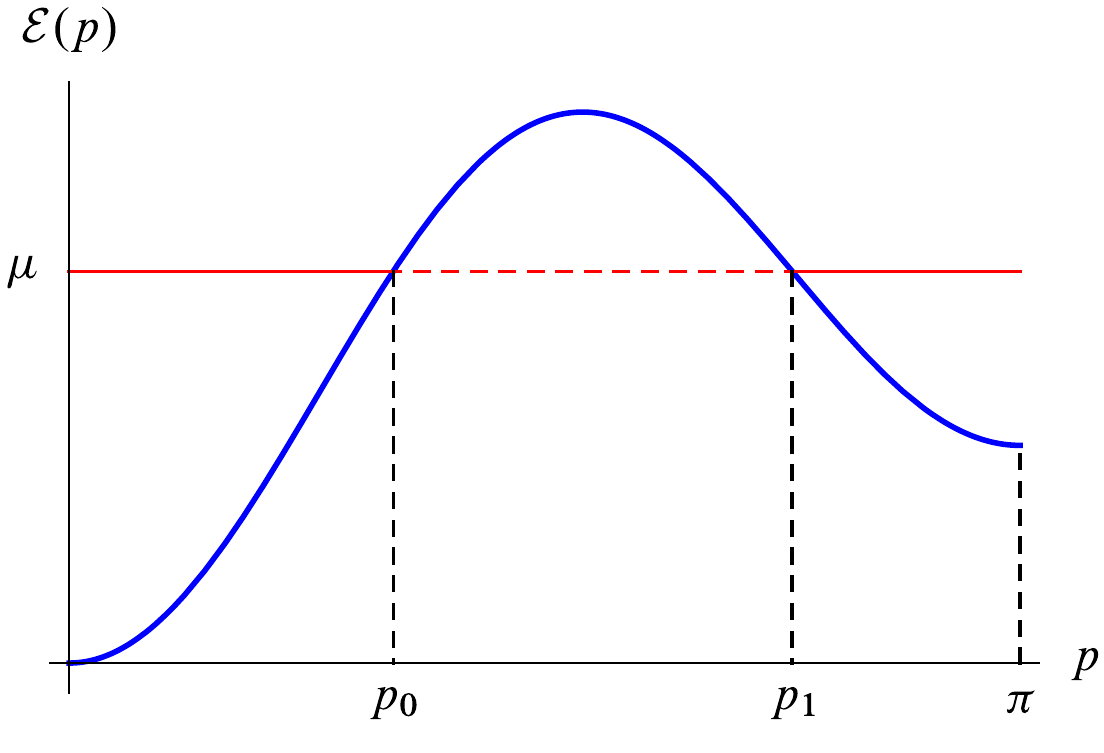}
  \caption{Dispersion relation~$\cE(p)$ with a single maximum in~$(0,\pi)$.}
  \label{fig.disp}
\end{figure}

As a second example, we shall consider a simple case in which the dispersion relation $\cE$ is not
monotonic in $[0,\pi]$. More precisely, suppose that $\cE$ is nonnegative and has a single
maximum in the open interval~$(0,\pi)$ (cf.~Fig.~\ref{fig.disp}).
For $\mu\in(0,\cE(\pi))$, the equation~$\cE(p)=\mu$ has one root~$p_0$ in the interval~$(0,\pi)$,
and the determinant~$D_L(\la)$ is approximately given by Eq.~\eqref{DLp0}. We shall next determine
the asymptotic behavior of~$D_L(\la)$ when $\mu\in(\cE(\pi),\cE_{\mathrm{max}})$, where~$\Emax$ is
the maximum value of~$\cE(p)$, and thus the equation~$\cE(p)=\mu$ has two roots~$p_0<p_1$ in the
interval~$(0,\pi)$. To begin with, from Eq.~\eqref{ALgen} it follows that in this case
\[
  (A_L)_{jk}=\frac1{2\pi}\left(\int_{-p_0}^{p_0}+\int_{p_1}^{2\pi-p_1}\right)\e^{-\iu(j-k)p}\diff
  p\,.
\]
Thus the symbol of~$T_L=\la+1-2A_L$ is given by
\begin{equation}\label{cp0p1}
  c\bigl(\e^{\iu\th}\bigr)=
  \begin{cases}
    \la-1\,,\quad & \th\in(-p_0,p_0)\cup(p_1,2\pi-p_1)\\
    \la+1\,,&\th\in(p_0,p_1)\cup(2\pi-p_1,2\pi-p_0)\,.
  \end{cases}
\end{equation}
In other words,~$c\bigl(\e^{\iu\th}\bigr)$ alternatively takes on the two values~$\la-1$
and~$\la+1$ on each of the four intervals~$(-p_0,p_0),\dots,(2\pi-p_1,2\pi-p_0)$ on
which~$\cE(p)-\mu$ has constant sign, starting with~$\la-1$. Since the symbol~\eqref{cp0p1} has
four discontinuities at the points~$\e^{\pm\iu p_0}$ and $\e^{\pm\iu p_1}$\,, we shall try to
express it as
\begin{multline}
  c\bigl(\e^{\iu\th}\bigr)=b\bigl(\e^{\iu\th}\bigr)t_{\be_0}\bigl(\e^{\iu(\th+p_0)}\bigr)t_{-\be_0}\bigl(\e^{\iu(\th-p_0)}\bigr)\\\times
  t_{\be_1}\bigl(\e^{\iu(\th+p_1)}\bigr)t_{-\be_1}\bigl(\e^{\iu(\th-p_1)}\bigr)
  \label{cp0p1can}
\end{multline}
for suitably chosen~$b$, $\be_i$. In fact, we only need
compute~$t_{\pm\be_1}\bigl(\e^{\iu(\th\pm p_1)}\bigr)$, which is straightforward:
\begin{align*}
  t_{\be_1}\bigl(\e^{\iu(\th+p_1)}\bigr)&=
                                          \begin{cases}
                                            \e^{\iu\be_1(\th+p_1-\pi)}\,,\en& -p_0<\th<2\pi-p_1\\
                                            \e^{\iu\be_1(\th+p_1-3\pi)}\,,& 2\pi-p_1<\th<2\pi-p_0
                                          \end{cases}\\
  t_{-\be_1}\bigl(\e^{\iu(\th-p_1)}\bigr)&=
                                           \begin{cases}
                                             \e^{-\iu\be_1(\th-p_1+\pi)}\,,\en& -p_0<\th<p_1\\
                                             \e^{-\iu\be_1(\th-p_1-\pi)}\,,& p_1<\th<2\pi-p_0\,.
                                           \end{cases}
\end{align*}
Combining the previous equations with Eq.~\eqref{tbp0} and comparing with Eq.~\eqref{cp0p1} we
immediately arrive at the system
\begin{align*}
  &b\,\e^{2\iu(\be_0p_0+\be_1p_1)}\e^{-2\pi\iu(\be_0+\be_1)}=b\,\e^{2\iu(\be_0p_0+\be_1p_1)}=\la-1\\
  &b\,\e^{2\iu(\be_0p_0+\be_1p_1)}\e^{-2\pi\iu\be_1}=\la+1\,.
\end{align*}
From the first equation it follows that $\be_0+\be_1$ must be an integer. Choosing the simplest
solution~$\be_0=-\be_1\equiv\be$ and dividing the last equation by the first one we again obtain
Eq.~\eqref{bebmono} for~$\be$. Finally, from the last equation it follows that
\begin{equation}\label{bpop1}
  b=(\la+1)\e^{2\iu\be(p_1-p_0-\pi)}=(\la+1)\bigg(\frac{\la+1}{\la-1}\bigg)^{(p_1-p_0-\pi)/\pi}\,.
\end{equation}
Note that~$\be$ is still given by Eq.~\eqref{bebmono}, so the condition~$|\Re\be|<1/2$, necessary
for the validity of the Fisher--Hartwig conjecture, also applies in this case if
$\la\notin[-1,1]$. Since now $M=-4\be^2$,
\[
  \prod_{r=1}^4G(1+\be_r)G(1-\be_r)=G(1+\be)^4G(1-\be)^4
\]
and
\begin{multline*}
  \prod_{1\le s<r\le 4}\big(2|\sin\bigl(\tfrac{\th_r-\th_s}2\bigr)|\big)^{2\be_r\be_s}\\
  = \big(2\sin\bigl(\tfrac{p_1-p_0}2\bigr)\big)^{-4\be^2}
  \big(2\sin\bigl(\tfrac{p_1+p_0}2\bigr)\big)^{4\be^2}\\
  \times(4\sin p_0\sin p_1)^{-2\be^2}\,.
\end{multline*}
By the Fisher--Hartwig conjecture, the determinant $D_L(\la)$ is given in this case by
\begin{align}
  D_L(\la)&=\bigg[\frac{4L^2\sin p_0\sin p_1
            \sin^2\bigl(\tfrac{p_1-p_0}2\bigr)}{\sin^2\bigl(\tfrac{p_1+p_0}2\bigr)}\bigg]^{\!\!-2\be^2}\notag\\
          &\hphantom{={}\bigg(}\times(\la+1)^L\bigg(\frac{\la+1}{\la-1}\bigg)^{\frac L\pi(p_1-p_0-\pi)}\notag\\
          &\hphantom{={}\bigg(}\times G(1+\be)^4G(1-\be)^4\big(1+o(1)\big)\,.
            \label{DLp0p1}
\end{align}

\section{Computation of the integral~$I_1(\al)$}\label{app.int}
In this appendix we shall provide an elementary derivation of the integral~$I_1(\al)$ in
Eq.~\eqref{I1}, which appears in the asymptotic expression of the Rényi entanglement entropy of
the model~\eqref{Hchain}. To begin with, we have
\[
  I_1(\al)=\frac2{\pi^2}(1-\al)^{-1}\hat I_1(\al)\,,
\]
with
\[
  \hat I_1(\al)=\int_{-1}^1\log\biggl[\bigg(\frac{1+x}{2}\bigg)^\al+
  \bigg(\frac{1-x}{2}\bigg)^\al\biggr]\,\frac{\diff x}{1-x^2}\,,
\]
or equivalently (performing the change of variables~$t=(x+1)/2$)
\begin{equation}\label{hI1al}
  \hat I_1(\al)=\int_0^1\log\bigl[t^\al+(1-t)^\al\bigr]\frac{\diff t}t\,.
\end{equation}
Integrating by parts we obtain the equivalent expression
\begin{equation}\label{hI1allog}
  \hat I_1(\al)=-\al\int_0^1\frac{t^{\al-1}-(1-t)^{\al-1}}{t^\al+(1-t)^\al}\,\log t\,\diff t\,.
\end{equation}
On the other hand, differentiating Eq.~\eqref{hI1al} with respect to~$\al$ we easily get
\[
  \hat I_1'(\al)=\int_0^1\frac{t^{\al-1}}{t^\al+(1-t)^\al}\,\frac{\log t}{1-t}\,\diff t\,,
\]
and hence~\cite{OLBC10}
\[
  \hat I_1'(\al)+\frac{\hat I_1(\al)}{\al}=\int_0^1\frac{\log t}{1-t}\,\diff t=-\frac{\pi^2}6\,.
\]
Solving this linear differential equation with the initial condition~$\hat I_1(1)=0$ we finally
obtain
\begin{equation}\label{hI1alfinal}
  \hat I_1(\al)=\frac{\pi^2}{12\al}\,(1-\al^2)\,,
\end{equation}
which immediately yields Eq.~\eqref{I1alfinal}.

\section{Simplification of the constant $\widetilde C_\al$}\label{app.Cal}

In this appendix we derive Eq.~\eqref{Calpsi} for the constant~$\widetilde C_\al$ in
Eq.~\eqref{Salfinal}, and show that it can be more simply expressed by means of the
integral~\eqref{Calsimp}. To this end, we use the elementary identity
\[
  \psi(z)=-\ga_E+\sum_{n=1}^\infty\bigg(\frac1{n}-\frac1{n+z-1}\bigg)\,,
\]
which can be immediately derived from the well-known infinite product for the Gamma function. Our
starting point is the definition~\eqref{I2} of~$I_2(\al)$, which can be written as
\begin{equation}\label{I2fz}
  I_2(\al)=
  \frac2{\pi^2}\int_{-1}^1\frac{s_\al(x)}{1-x^2}\big[f\bigl(Z(x)\bigr)+f\bigl(1-Z(x)\bigr)\big] \diff x\,,
\end{equation}
with $Z(x)=1/2+\iu B(x)$ and
\[
  f(z)=\sum_{n=1}^\infty\frac{z^3}{n(n^2-z^2)}\,.
\]
From the relation
\[
  \frac{z^2}{n(z^2-n^2)}=\frac12\bigg(\frac1{n+z}-\frac1n\bigg)
  +\frac12\bigg(\frac1{n-z}-\frac1n\bigg)
\]
and the functional identity $\psi(z+1)=\psi(z)+1/z$ satisfied by the digamma function it
immediately follows that
\[
  f(z)=-\frac z2\,\big[\psi(z)+\psi(1-z)\big]-z\ga_E-\frac12\,,
\]
and therefore
\[
  f(z)+f(1-z)=-1-\ga_E-\frac12\,\big[\psi(z)+\psi(1-z)\big]\,.
\]
Substituting into Eq.~\eqref{I2fz} and using Eqs.~\eqref{I1alfinal} and~\eqref{Calfinal} we easily
obtain Eq.~\eqref{Calpsi}.

In order to prove Eq.~\eqref{Calsimp}, we first make the change of variable $w=B(x)$ in
Eq.~\eqref{Calpsi}, which yields
\begin{align*}
  &(1-\al)\widetilde C_\al=-\frac2\pi\int_0^\infty\log\biggl[\frac{2\cosh(\pi\al w)}{\big(2\cosh(\pi
    w)\big)^\al}\biggr]\\&\kern9em\times\big[\psi\bigl(\tfrac12+\iu\,w\bigr)
                           +\psi\bigl(\tfrac12-\iu\,w\bigr)\big]\,\diff w\\
  &=\frac2{\iu\pi}\int_0^\infty\log\biggl[\frac{2\cosh(\pi\al w)}{\big(2\cosh(\pi
    w)\big)^\al}\biggr]\frac{\diff}{\diff w}\log\bigg[
    \frac{\Ga\bigl(\tfrac12-\iu\,w\bigr)}{\Ga\bigl(\tfrac12+\iu\,w\bigr)}\bigg]\,\diff w\,.
\end{align*}
We next integrate by parts, taking into account that by Stirling's formula we have
\[
  \log\bigg[ \frac{\Ga\bigl(\tfrac12-\iu\,w\bigr)}{\Ga\bigl(\tfrac12+\iu\,w\bigr)}\bigg] =O(w\log
  w)
\]
while
\[
  \log\biggl[\frac{2\cosh(\pi\al w)}{\big(2\cosh(\pi
    w)\big)^\al}\biggr]=O\bigl(\e^{-2\pi\min(1,\al)w}\bigr)\,,
\]
so that the boundary term vanishes. We thus obtain
\begin{multline*}
  (1-\al)\widetilde C_\al\\=2\iu\al\int_0^\infty\!\big[\tanh(\pi\al w)-\tanh(\pi w)\big]\log\bigg[
  \frac{\Ga\bigl(\tfrac12-\iu\,w\bigr)}{\Ga\bigl(\tfrac12+\iu\,w\bigr)}\bigg]\diff w\,.
\end{multline*}
On the other hand, from Gauss's integral representation of the digamma function~\cite{WW27}
\[
  \psi(z)=\int_0^{\infty}\bigg(\frac{\e^{-t}}{t}-\frac{\e^{-zt}}{1-\e^{-t}}\bigg)\,\diff t
\]
it easily follows that
\[
  \log\Gamma(z)=\int_0^\infty\bigg[z-1-\frac{1-\e^{-(z-1)t}}{1-\e^{-t}}\bigg]\frac{\e^{-t}}t\,\diff
  t
\]
and hence
\[
  \log\bigg[ \frac{\Ga\bigl(\tfrac12-\iu\,w\bigr)}{\Ga\bigl(\tfrac12+\iu\,w\bigr)}\bigg]
  =\iu\int_0^\infty\bigg[\csch(t/2)\sin(wt)-2w\e^{-t}\bigg]\frac{\diff t}t\,.
\]
Substituting into the last formula for~$\widetilde C_\al$ and using the elementary identity
\[
  \tanh(\pi\al w)-\tanh(\pi w)=\frac{2\e^{-2\pi w}}{1+\e^{-2\pi w}} -\frac{2\e^{-2\pi\al
      w}}{1+\e^{-2\pi\al w}}
\]
we obtain
\[
  (1-\al)\widetilde C_\al=4\al\int_0^\infty\frac{g_1(t)-g_\al(t)}t\,\diff t\,,
\]
with
\[
  g_\al(t)=\int_0^\infty\frac{\e^{-2\pi\al w}}{1+\e^{-2\pi\al
      w}}\big[2w\e^{-t}-\csch(t/2)\sin(wt)\big]\diff w.
\]
The latter integral can be evaluated in closed form by elementary means. Indeed,
\begin{align*}
  &\int_0^\infty\frac{w\,\e^{-2\pi\al w}}{1+\e^{-2\pi\al
    w}}\diff w=\sum_{n=1}^\infty(-1)^{n+1}\int_0^\infty w\,\e^{-2n\pi\al w}\diff w\\
  &=\frac1{4\pi^2\al^2}\sum_{n=1}^\infty\frac{(-1)^{n+1}}{n^2}=\frac{\ze(2)}{8\pi^2\al^2}
    =\frac1{48\al^2}\,,
\end{align*}
while
\begin{align*}
  &\int_0^\infty\frac{\e^{\iu wt}\,\e^{-2\pi\al w}}{1+\e^{-2\pi\al
    w}}\diff w=\sum_{n=1}^\infty(-1)^{n+1}\int_0^\infty\e^{(\iu t-2n\pi\al) w}\diff w\\
  &=\sum_{n=1}^\infty\frac{(-1)^n}{\iu\,t-2n\pi\al}\,,
\end{align*}
and therefore
\begin{align*}
  &\int_0^\infty\frac{\e^{-2\pi\al w}\sin(wt)}{1+\e^{-2\pi\al
    w}}\diff w=\Im\sum_{n=1}^\infty\frac{(-1)^n}{\iu\,t-2n\pi\al}\\
  &=t\sum_{n=1}^\infty\frac{(-1)^{n+1}}{t^2+4\al^2\pi^2n^2}
    =\frac1{4\al}\bigg(\frac{2\al}t-\csch\bigl(t/(2\al)\bigr)\bigg).
\end{align*}
We thus obtain
\[
  g_\al(t)=\frac{\e^{-t}}{24\al^2}+\csch(t/2)\bigg(\frac{\csch\bigl(t/(2\al)\bigr)}{4\al}
  -\frac1{2t}\bigg)\,,
\]
from which Eq.~\eqref{Calsimp} easily follows.

% \onecolumngrid
% \bibliographystyle{apsrev4-1}
% \bibliography{cmprefs}

\end{document}